\documentclass[twocolumn]{aastex7}

\newcommand{\mathbfit}[1]{\textbf{\textit{#1}}}

\usepackage{amsmath}

\begin{document}

\title{Modeling YSO Jets in 3D II: Accretion-Fed, Star-Anchored Poynting Jets in the Low-Density Polar Cavity Powered by Disk-Magnetosphere Interaction}

\author[0000-0003-2929-1502]{Yisheng Tu}
\affiliation{Astronomy Department, University of Virginia, Charlottesville, VA 22904, USA}
\affiliation{Virginia Institute of Theoretical Astronomy, University of Virginia, Charlottesville, VA 22904, USA}
\email[show]{yt2cr@virginia.edu}  

\author{Zhi-Yun Li}
\affiliation{Astronomy Department, University of Virginia, Charlottesville, VA 22904, USA}
\affiliation{Virginia Institute of Theoretical Astronomy, University of Virginia, Charlottesville, VA 22904, USA}
\email{zl4h@virginia.edu}

\author{Zhaohuan Zhu}
\affiliation{Department of Physics and Astronomy, University of Nevada, Las Vegas, NV, 89154-4002, USA}
\affiliation{Nevada Center for Astrophysics, University of Nevada, Las Vegas, 4505 S. Maryland Pkwy, Las Vegas, NV, 89154, USA}
\email{zhaohuan.zhu@unlv.edu}

\author{Xiao Hu}
\affiliation{Astronomy Department, University of Virginia, Charlottesville, VA 22904, USA}
\affiliation{Department of Astronomy, University of Florida, Gainesville, FL 32608, USA}
\email{xiao.hu.astro@gmail.com}

\author{Chun-Yen Hsu}
\affiliation{Astronomy Department, University of Virginia, Charlottesville, VA 22904, USA}
\affiliation{Virginia Institute of Theoretical Astronomy, University of Virginia, Charlottesville, VA 22904, USA}
\email{kdj8qp@virginia.edu}

\begin{abstract}
The origin of jets in young stellar objects (YSOs) remains a subject of active investigation. We present a 3D magnetohydrodynamic simulation of jet launching in YSOs, focusing on the interaction between the stellar magnetosphere and the accretion disk. In our model, a fast, low-density bipolar jet is powered by disk-magnetosphere interaction and launched through the polar cavity that is mass-loaded from the disk rather than the star. Specifically, outflows are driven by toroidal magnetic pressure generated along “two-legged” field lines, anchored at a magnetically dominated stellar footpoint and a mass-dominated point on the (magnetically elevated) disk surface via a cyclic “load–fire–reload” process: in the “load” stage, differential rotation between stellar and disk footpoints generates toroidal magnetic pressure; in the “fire” stage, vertical gradients in the toroidal field accelerate plasma and transport Poynting flux into the polar cavity; in the “reload” stage, magnetic reconnection allows the cycle to repeat, reforming “two-legged” field lines. These field lines are not required to be fully reset to a dipolar loop configuration; it is only required that the disk-end be shallowly embedded in the (elevated) disk surface. This rapid, asynchronous process produces a continuous, large-scale outflow. The resulting magnetically dominated (Poynting) jet, accelerated by magnetic pressure within the low-density polar cavity, is distinct from the denser, slower disk wind launched 
through the classic magnetic-tower mechanism. Comparison with a disk-only model shows that the rotating stellar magnetosphere promotes {\it bipolar} jet launching by shaping a magnetic geometry favorable to symmetric outflows.
\end{abstract}


\section{Introduction}

Star formation is often viewed as a collapse process, where a dense core, typically several to tens of thousands of AUs in size, collapses to form a young stellar object (YSO) consisting of a central star, a circumstellar disk, and, ultimately, planets. However, the reverse process, outflow, is also a fundamental aspect of the formation and evolution of YSOs. Outflows play a crucial role in shaping the chemical and dust distribution within individual star-forming systems \citep[][]{Tsukamoto2021, Cacciapuoti2024, Morbidelli2024} and influence the formation of nearby stars and planets by regulating the dynamics of molecular clouds \citep[][]{Andre2014, Guszejnov2021, Hsieh2023}.

The most energetic component of the outflow is the jet, a highly collimated stream of gas traveling at speeds exceeding $10^7~\mathrm{cm~s}^{-1}$, often extending thousands of AU or more in length \citep[][]{Podio2015, Ray2021}. Recent observations have provided greater insight into the structure of jets and their surrounding environments \citep[e.g.,][]{Codella2014, deValon2020, Lee2021, Whelan2024, Assani2024, Lopezva2024, Hsu2024, Delabrosse2024}. Accompanying the jet is a slower, more massive outflow known as the disk wind, which typically surrounds the jet with a wider opening angle \citep[][]{Tabone2017, Pascucci2023}. The difference in the properties of the jet and disk wind suggests these two outflows are launched from different locations and likely with different magnetic lever arms if they are driven magnetically \citep[][]{Lee2017,deValon2020, Delabrosse2024}. Since both the jet and disk wind significantly impact both local and global star formation processes, understanding their driving mechanisms is crucial to understanding star formation. A key open question is whether the jet and disk wind share a common driving mechanism, particularly what powers the jet itself.

It is becoming a consensus that the jets and outflows are most likely magnetically driven. One commonly used model is the class of magnetocentrifugal models \citep[][]{Shu1995, Ferreira1997, Krasnopolsky1999, Ustyugova1999, Krasnopolsky2003, Anderson2005, Anderson2006, Bai2013, Jacquemin-Ide2021}, where the centrifugal force launches outflows as a result of angular momentum transportation along a magnetic field line. While the magneto-centrifugal model and its variants, including the ``disk wind'' \citep[][]{Blandford1982, Pudritz1983} and ``X-wind'' \citep[][]{Shu1994, Shu2000, Shang2002, Cai2008} models, can effectively explain the excessive angular momentum carried by the jet, how the exact launching conditions can be met under observational and theoretical constraints remains unclear \citep[][]{Desch2010, Bjerkeli2016}.

Another class of models is the Poynting flux-driven jet model, where the outflow is launched by a predominantly toroidal magnetic field that transfers energy and angular momentum to the gas \citep[][]{Shibata1986, Lovelace2001, Lovelace2009, Guan2014}. A related model is the ``magnetic-tower" model \citep[][]{Lynden1996, Lynden2003, Kato2004, Nakamura2007}, where the loop-like toroidal magnetic fields stack on top of each other, forming a magnetic loop tower. This class of models is usually more mass-loaded than the magneto-centrifugal model \citep[][]{Jacquemin-Ide2019}, resulting in a more gradual acceleration. Although this class of models provides an alternative method of acceleration, the stability of the required magnetic condition and the external pressure necessary to collimate the jet pose challenges to this model \citep[][]{Moll2008, Huarte-Espinosa2012}.

We should note that, at the deepest level, the classic magneto-centrifugal and magnetic tower models are fundamentally the same, as discussed in, e.g., \citet{Ferreira1997} and \citet[][see his Sec. 3.2]{Spruit2010}.
Either launching mechanism relies on energy and angular momentum extraction from its base, which is typically believed to be rooted in the inner disk or the star. One possibility is that the jet is driven by the disk alone. Since the YSO disk is expected to be magnetized, particularly during the early phase of disk evolution \citep[][]{Wurster2018, Vlemmings2019, Galametz2020, Lankhaar2022, Yen2024, Tu2024a, Ohashi2025}, the magnetic field in the disk could channel the disk's rotational energy and momentum to a large distance, forming a jet. 
Previous studies have investigated this possibility \citep[e.g.][]{Casse2002, Kato2002, Anderson2006, Zanni2007, Tzeferacos2009, Fendt2011, Murphy2010, Fendt2013, Stepanovs2016, Bethune2017, Zhu2018, Ramsey2019, Mattia2020a, Mattia2020b, Mishra2020, Jacquemin-Ide2021, Jannaud2023, Takaishi2024, Tu2025a, Tu2025b, Takasao2025}. However, when a well-ionized, relatively weakly magnetized inner disk is included self-consistently in our previous 3D global jet-formation simulations without a prescribed magnetic diffusivity, we found that the jet was relatively weak and often unipolar. It motivates us to include in the present study a key ingredient missing from our previous work: the stellar magnetosphere.

Observations have shown that YSOs could have a strong magnetosphere, and previous studies and simulations have mainly focused on how the magnetosphere influences the disk accreting onto the star \citep[][]{Elsner1977, Bardou1996, Miller1997, Bessolaz2008, Zanni2009, Romanova2009, Romanova2012, Zanni2013, Lai2014, Takasao2019, Ireland2021, Romanova2021, Romanova2025, Zhu2024, Zhu2025}. Outflow launching from the magnetosphere-disk boundary has been examined primarily in 2D \citep[e.g.,][]{Goodson1999, Romanova2005, Lovelace2013, Cemeljic2013, Lovelace2014}, and less in 3D \citep[e.g.,][]{Romanova2009, Takasao2022, Zhu2025}, showing the possibility of jet launching. Yet, how the magnetosphere participates in jet launching and how it interacts with the disk field remains uncertain. This paper focuses on modeling these key processes in a self-consistent 3D non-ideal magneto-hydrodynamic global model.

This study extends the work of \citet{Tu2025a} and \citet{Tu2025b}, who investigated outflow launching using the disk gas and the disk magnetic field alone. \citet{Tu2025a} performed a 2D axisymmetric model and identified a key ingredient in jet launching, the ``avalanche accretion stream" along the disk surface, which actively provides the energy and momentum to the jet-launching gas.  \citet{Tu2025b} extended the models to 3D and identified a light-loaded magnetic pressure-driven jet model in the disk-only case, whose jet is highly variable and asymmetric in both disk hemispheres\footnote{ We note that \citet{Zhu2018} and \citet{Jacquemin-Ide2021} also performed disk-only simulations. Some similarities and differences between these works and ours are discussed in \citet{Tu2025a} (see their Section 4.1 in particular).}. In this study, we follow the disk setup in \citet{Tu2025b} and include the stellar magnetosphere to examine the effect of the magnetosphere on jet launching.

This paper is organized as follows. Sec.~\ref{sec:method} describes the simulation setup and model parameters. Sec.~\ref{sec:result} gives an overview of the simulation result and identifies the key components of the modeled jet-launching system. The jet-launching mechanism is described in detail in Sec.~\ref{sec:jet-launching-mechanism}, followed by a discussion of this mechanism in the context of the literature in Sec.~\ref{sec:discussion}. We conclude in Sec.~\ref{sec:conclusions}.

\section{Method}
\label{sec:method}

The following set of non-ideal MHD equations governs the evolution of the circumstellar disk threaded by open magnetic field lines with the inner disk truncated by the stellar magnetosphere:
\begin{equation}
    \frac{\partial\rho}{\partial t} + \nabla\cdot(\rho\mathbfit{v}) = 0,
\end{equation}
\begin{equation}
    \rho\frac{\partial \mathbfit{v}}{\partial t} + \rho(\mathbfit{v}\cdot\nabla)\mathbfit{v} = -\nabla P + \frac{1}{c}\mathbfit{J}\times\mathbfit{B} - \rho\mathbfit{g},
    \label{equ:mhd momentum equation}
\end{equation}
\begin{equation}
    \frac{\partial \mathbfit{B}}{\partial t} = \nabla\times(\mathbfit{v}\times\mathbfit{B}) - \frac{4\pi}{c}\nabla\times[\eta_O(\mathbfit{J}\times\hat{\mathbfit{B}})],
    \label{equ:mhd induction}
\end{equation}
where $\mathbfit{J} = (c/4\pi)\nabla\times\mathbfit{B}$ is the current density, and $\eta_O$ the Ohmic dissipation coefficient. Other symbols have their usual meanings. 

We use the \texttt{Athena++} code \citep{Stone2020} to solve the governing equations in Cartesian coordinates with static mesh refinement (SMR). The initial condition setup is similar to \citet{Tu2025b}, with the addition of the stellar magnetosphere. We refer the readers to \citet{Tu2025b} for a detailed description of the hydro, ionization, and non-ideal MHD setup, which are briefly summarized here.  

We adopt a power law $\rho = \rho_0 (r/r_0)^{-1.35}$ for the density of the disk midplane with $\rho_0 = 10^{-9}\mathrm{\ g\ cm^{-3}}$ and $r_0 = 6.3\times10^{-2}\ \mathrm{au}$ following \citet{Tu2025a}. To model the truncation of the disk due to the stellar magnetosphere, the midplane density decreases as
\begin{equation}
    \rho(r < r_\mathrm{trun}) = \rho \Big(\frac{r}{r_\mathrm{trun}}\Big)^p,
\end{equation}
where $r_\mathrm{trun} = 0.1$~au and $p = -10$. The remainder of the density is prescribed by assuming the disk is in isothermal hydrostatic equilibrium. To minimize numerical problems and ensure runability, 
we implement a similar density floor as in \citet{Tu2025a}:
\begin{equation}
    \rho_\mathrm{floor}(r) = 
    \begin{cases}
      \rho_f ;             & r < r_f, \\
      \rho_f(r / r_f)^{q_f}; & \text{otherwise},
    \end{cases}
    \label{equ:dfloor}
\end{equation}
where $\rho_f = 6\times10^{-12}$~g cm$^{-3}$, $r_f = 0.015$~au, and $r$ is the spherical radius. $q_f = -5$ allows a rapid drop in the density floor towards a large radius where a density floor is unnecessary. To speed up the simulation during the later stages, we increase the density floor coefficient $\rho_f$ to $2.4\times10^{-11}$~g cm$^{-3}$ after $t=1.0$~yr. The thermal and non-ideal MHD setup is identical to \citet{Tu2025b}, where thermal ionization of alkali metals keeps the inner disk ($r\lesssim0.13$~au, the active zone) and the envelope in the ideal MHD limit, and Ohmic dissipation dominates the outer disk ($r\gtrsim0.13$~au, the dead zone).

To quantify and trace the amount of mass introduced by the density floor, we utilize the ``passive scalars'' in \texttt{Athena++} to differentiate between the real gas in the simulation (defined as all the mass present in the active domain at initialization) and the added gas introduced by the density floor. The real gas dominates the outflow and outflow-launching region throughout the simulation. In contrast, the added mass primarily resides in the low-altitude infalling region around the polar axis (i.e., the polar region shown in Figure~\ref{fig:starfield}) and the inner magnetosphere, where the star rapidly accretes it.

Following \citet{Zhu2024} and \citet{Zhu2025}, we incorporate a dense ``star'' of $1~M_\odot$ at the center of the simulation to minimize numerical diffusion of the stellar magnetosphere. This approach leverages the Cartesian coordinate system, which avoids the presence of voids within the simulation domain. The stellar radius is $R_\star = 0.015~\mathrm{au}$ (about $3.2~R_\odot$), with a central density of $1.4~\mathrm{g~cm^{-3}}$. The remainder of the stellar density is calculated assuming the star is in isothermal hydrostatic equilibrium. To avoid numerical problems due to too large a density gradient at the stellar surface, we separate the star from the remainder of the simulation domain by setting a ``fixing radius'' $r_\mathrm{fix} = 0.018$~au, within which the hydro quantities are reset to their initial values at each time step, while the magnetic field remains untouched. The passive scalars are set to 0 within $r_\mathrm{fix}$ at every time step. The gas in $r_\mathrm{fix}$ is assumed to rotate as a solid body with a rotational period of $12~\mathrm{days}$ (corresponding to the Keplerian rotation period at $0.1~\mathrm{au}$). 

The gravitational force is modeled by assuming that stellar gravity dominates the gravitational acceleration everywhere in the simulation. The stellar gravity is modeled as a point-source gravity, with a smoothing towards $r = 0$ to avoid the singularity
\begin{equation}
    a_g(r) =
    \begin{cases}
      0;             & r < r_s, \\
      -\frac{GM_\star}{r^2}\frac{(r - r_s)^2}{(r - r_s)^2 + r_s^2}\hat{\mathbfit{r}}; & \text{otherwise},
    \end{cases}
\end{equation}
where $r_s = 0.012~\mathrm{au}$. 

To ensure divergence-free initial condition and model both the open magnetic field lines threading the disk and the stellar magnetosphere, the global vector potential is the sum of the vector potential of the magnetosphere and the open disk field, i.e.
\begin{equation}
    \mathbfit{A} = \mathbfit{A}_\mathrm{disk} + \mathbfit{A}_\mathrm{mag},
\end{equation}
where $\mathbfit{A}_\mathrm{disk}$ is the same as the vector potential used in \citet{Tu2025b}, where the disk is threaded by an open $+\hat{z}$-direction magnetic field.
The stellar magnetosphere is modeled following \citet{Zhu2024}, using a vector potential
\begin{equation}
    \mathbfit{A}_\mathrm{mag} = \frac{\Bar{\mathbfit{m}}\times\mathbfit{r}}{r_c^3},
    \label{equ:Amag} 
\end{equation}
where $r_c = \max(r, r_s / 2)$. $\Bar{\mathbfit{m}}$ is chosen to be in the $-\hat{z}$ direction and of such magnitude that the initial magnetic field strength at the surface of the star is about 2000 Gauss. This choice of $\Bar{\mathbfit{m}}$ ensures that the magnetic field of the magnetosphere points upward on the disk midplane, aligned with the direction of the initial open disk field.

The computation domain and mesh refinement are chosen to best resolve the disk and the outflow-launching polar region while keeping a reasonable computational cost. The simulation domain spans $\pm~10~\mathrm{au}$ in the $\hat{x}$ and $\hat{y}$ directions, and $\pm~20~\mathrm{au}$ in the $\hat{z}$ direction. The root grid has a resolution of $64\times 64\times 128$, ensuring all cells are cubic. Nine levels of refinements are applied in the simulation, with the finest level applied in the disk ($\pm 0.04~\mathrm{au}$ in the $\hat{x}$ and $\hat{y}$ directions, and $\pm 0.02~\mathrm{au}$ in the $\hat{z}$ direction), and in the outflow region ($\pm 0.02~\mathrm{au}$ in the $\hat{x}$ and $\hat{y}$ directions, and $\pm 0.04~\mathrm{au}$ in the $\hat{z}$ direction). The remainder of the grid is generated automatically by applying the condition that adjacent meshblocks can only differ by at most 1 level of refinement. To reduce numerical diffusion within the star but not add too much computational cost, we apply two additional levels of refinement inside the star: level 10 covers $\pm 0.01~\mathrm{AU}$, and level 11 resolves the central $\pm 0.005~\mathrm{au}$. This refinement strategy keeps the stellar magnetosphere stable throughout the simulation. As a result of the mesh refinement, the finest cell size is $(0.0006~\mathrm{au})^3$ in the active domain, and $(0.00015~\mathrm{au})^3$ inside the star.

\section{Results}
\label{sec:result}
In this section, we present an overview of our simulation results and highlight key features of the jet structure. Because of the dynamic nature of the outflow launched in our model, we crudely divide the jet and disk wind using a simple velocity threshold: the jet is defined as gas with velocity $>100$ km s$^{-1}$ away from the disk, and the disk wind as gas between $10$ and $100$ km s$^{-1}$. The choice of $ 100$ km/s is somewhat arbitrary. Nevertheless, it separates well the low-density, fast-moving flow in the polar region from the much denser, more slowly moving surrounding flow in our particular simulation (see, e.g., Fig.~\ref{fig:diskwind}a and Fig.~\ref{fig:Br_flux}, below). We have verified that the bulk of the jet material defined this way is super-fast-magnetosonic, especially at relatively large distances from the central star (see Fig.\ref{fig:diskwind}a,b below), so that its launching should not be strongly modified by the jet interaction with its ambient medium.

\begin{figure*}
    \centering
    \includegraphics[width=\linewidth]{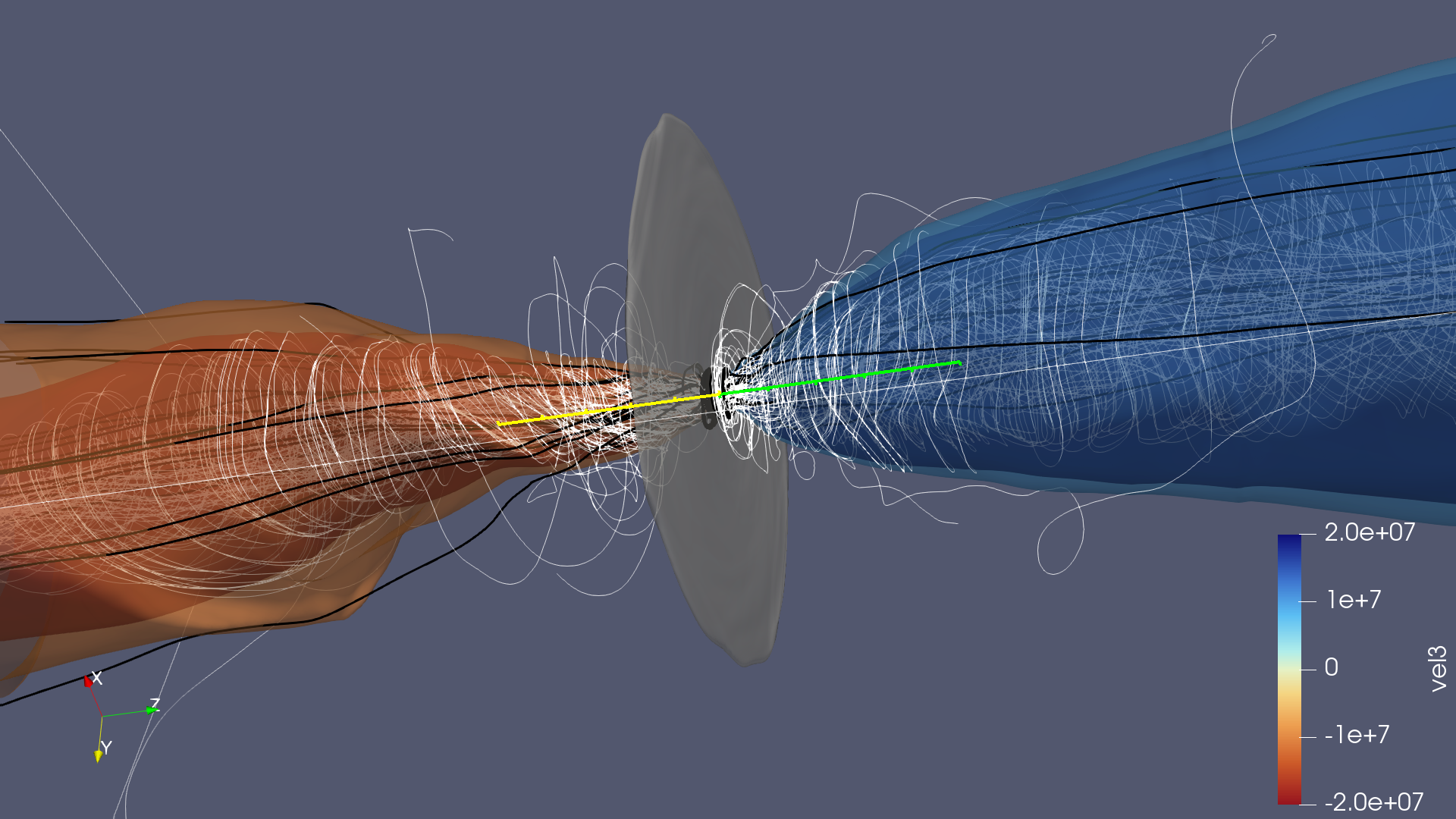}
    \caption{3D rendition of the bipolar jet launched in our model. The outer red and blue shells are isosurface contours at $100$ km s$^{-1}$, whereas the inner ones are at 200 km s$^{-1}$. The gray structure on the equatorial plane represents the disk. For reference, the green and yellow lines each have a length of 5 au, illustrating the physical scale of the system. An animated version of this figure can be found at: \url{https://figshare.com/s/67335cb1846eca1edcb1}. 
    The movie is 27 seconds long, highlighting the strong bipolar outflow from the simulation's beginning (0 yr) to the end (5.5 yr).}
    \label{fig:3D}
\end{figure*}

Figure~\ref{fig:3D} shows a 3D rendering of the jet in our model, where blue and red contours represent the outflow in the upper and lower hemispheres, respectively. Each hemisphere contains two layers: the outer layer encloses gas moving at speeds exceeding $\pm 100$ km s$^{-1}$, while the inner layer highlights gas with speeds exceeding $\pm 200$ km s$^{-1}$, illustrating the presence of a powerful bipolar jet. To avoid obscuring the jet, the disk is visualized as a toroidal structure around the midplane, with most of its mass removed for clarity. The white lines represent magnetic fieldlines, and velocity streamlines are shown by the black lines. 

The partial opening of the magnetosphere during the early stages of simulations involving a stellar dipole is a well-known process \citep[e.g.][]{Goodson1999, Romanova2009, Zanni2013, Lii2014, Ireland2021}. In this phase, the initially closed outer dipole field lines that thread the disk inflate and eject poloidal magnetic loops. This magnetic ejection results in the partial opening of the magnetosphere as the system undergoes its initial adjustment.

\begin{figure*}
    \centering
    \includegraphics[width=0.40\linewidth]{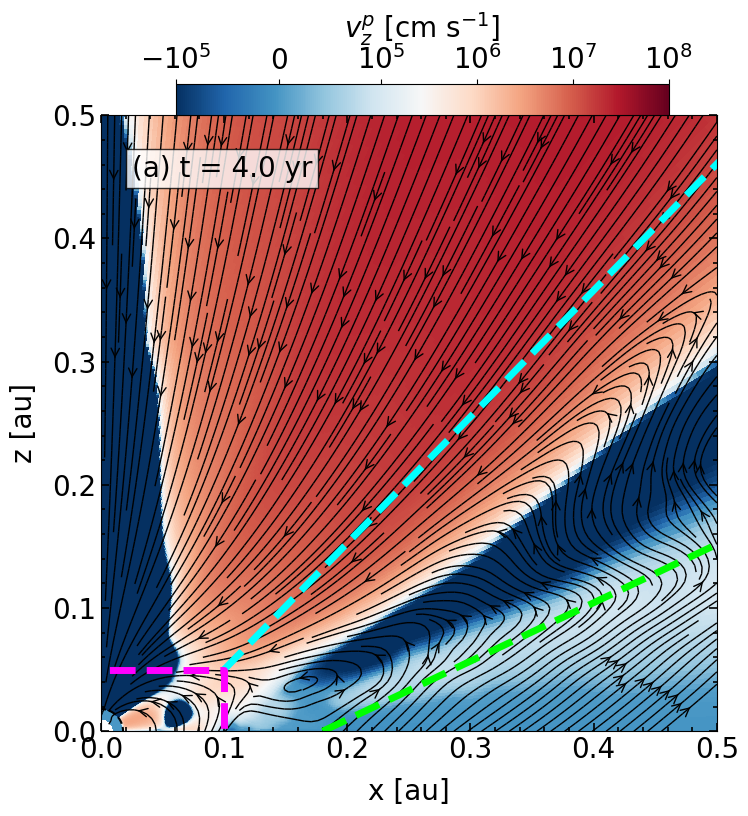}
    \includegraphics[width=0.59\linewidth]{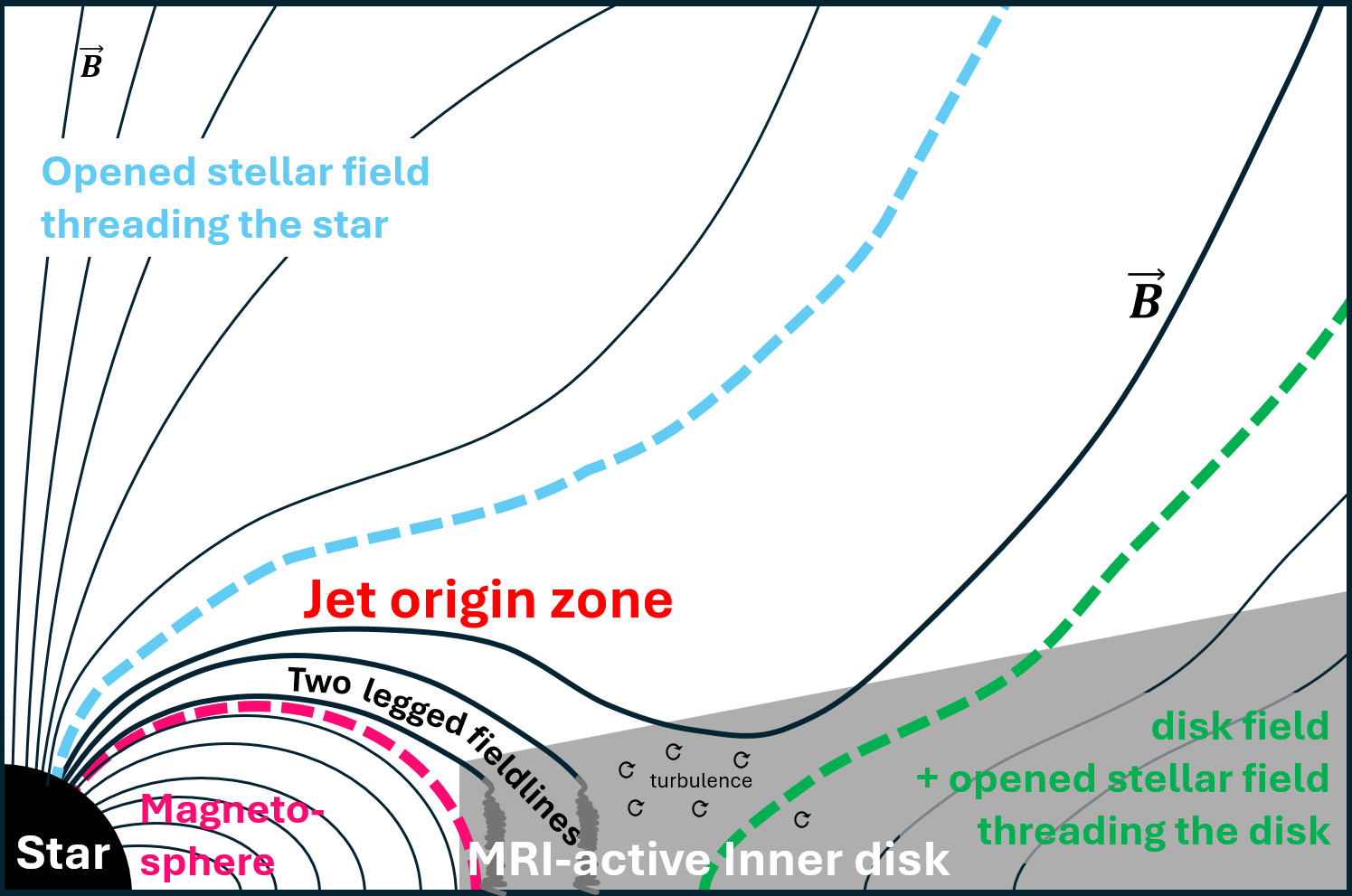}
    \caption{\textbf{Left panel:} Magnetic field geometry, represented by the time- and azimuthal-averaged magnetic field lines during the later stages of the simulation. The background color map shows the projected $\hat{z}$-direction velocity $v_z^p$ (equ.~\ref{equ:vzp}). \textbf{Right panel:} a cartoon illustration of the magnetic field geometry, and key components in the simulation. \textbf{Both panels:} The magnetic field structure can be roughly divided into four distinct zones, delineated by corresponding magenta, cyan, and green lines in both the left and right panels.  The magenta line encloses the permanently closed stellar magnetosphere; the cyan line marks the boundary of opened stellar field lines that still thread the star; the green line outlines the extent of the large-scale magnetic field lines threading the disk. The region between the cyan and green lines corresponds to a reconnecting, outflow-driving zone where the interaction between stellar and disk fields powers the outflow.  
    }
    \label{fig:init_open}
\end{figure*}

The initial adjustment results in a characteristic magnetic field geometry. To illustrate the magnetic field geometry and its relationship with the outflow, we show in Figure~\ref {fig:init_open} the time- and azimuthal-averaged magnetic fieldlines overplotted on the projected $\hat{z}$-direction velocity $v_z^p$, defined as
\begin{equation}
    v_z^p = v_z \frac{z}{|z|},
    \label{equ:vzp}
\end{equation}
so outflow in both the upper and lower hemispheres is positive whereas inflow is negative. Since the magnetic field connecting the star and the disk is essential in driving the outflow, we provide a sketch of the key elements involved in jet-launching in Figure~\ref{fig:init_open}(b).
The magnetic field geometry can be crudely divided into four zones: the permanently closed stellar magnetosphere near the star (enclosed by the magenta line in Figure~\ref{fig:init_open}), the $-\hat{z}$ direction opened stellar field attached to the star (roughly enclosed by the cyan lines), the large-scale open $+\hat{z}$-direction field (similar to the original disk field, the outer region marked by the green lines), and the less ordered field lines lying in between the former two (between the green and cyan lines). This overall magnetic field structure remains consistent throughout the simulation after the initial adjustment. Since the magnetic field configuration plays a crucial role in launching the outflow, we first examine its properties before discussing each outflow component in detail.

\subsection{Magnetic flux evolution}
\label{sec:mag_flux_evol}

\begin{figure*}
    \centering
    \includegraphics[width=\linewidth]{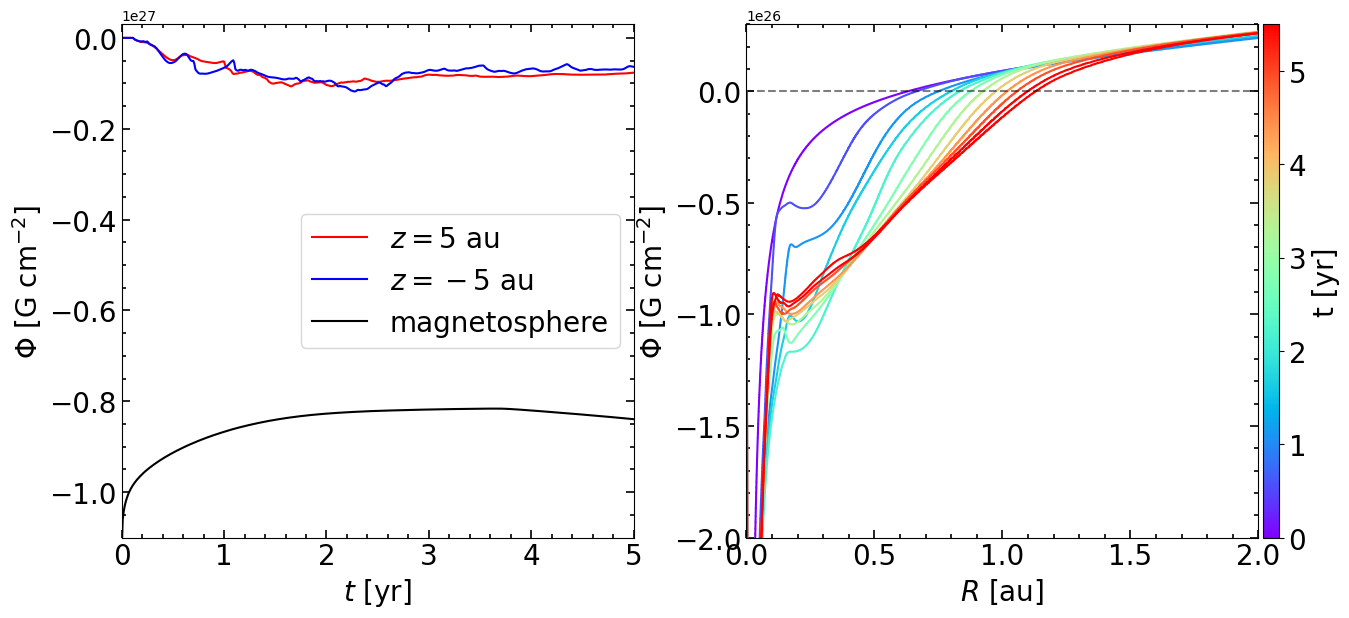}
    \caption{Magnetic flux evolution in the simulation. \textbf{The left panel} shows the total magnetospheric flux (black line), defined as the maximum magnitude of the total downward $\hat{z}$-directed magnetic flux integrated azimuthally at each cylindrical radius. The red and blue lines indicate the amount of opened stellar magnetosphere flux, measured as the maximum magnitude of azimuthally-integrated downward $B_z$ at $z=\pm5$ au above and below the disk midplane, respectively. \textbf{The right panel} shows the time evolution of the enclosed magnetic flux at each cylindrical radius at the midplane. The stellar magnetosphere primarily contributes to the negative flux at small radii, while the disk field dominates the positive flux at larger radii.}
    \label{fig:magnetic_flux}
\end{figure*}

To quantify the opening of the magnetosphere field during simulation, Figure~\ref{fig:magnetic_flux} shows the downward magnetic flux ($-\hat{z}$ direction) at $z = \pm5~\mathrm{au}$ as a function of time for the upper and lower hemispheres, respectively. Since the disk’s magnetic field is oriented upward, any downward flux at these heights must originate from the opening of the stellar magnetosphere. For reference, Figure~\ref{fig:magnetic_flux} also shows the total stellar magnetospheric flux, defined as the total downward magnetic flux at the midplane. Due to magnetic reconnection of the innermost magnetospheric loops inside the star, the total stellar magnetic flux decreases slightly at the beginning of the simulation before stabilizing at approximately $-8\times10^{26}~\mathrm{G\ cm^2}$. The opened magnetospheric flux remains around $-8\times10^{25}~\mathrm{G\ cm^2}$, accounting for about 10\% of the total magnetospheric flux.  

The opened magnetic flux dominates the magnetic field of the inner disk. It is equivalent to the total initial flux of the disk within $r\approx7~\mathrm {au}$, and diffuses outward gradually. Figure~\ref{fig:magnetic_flux} illustrates the evolution of the net magnetic flux enclosed within a given radius on the midplane over time. Due to the strong, negative magnetic field inside the star, the enclosed flux initially drops to a highly negative value at small radii, then gradually returns to positive as more upward-directed field lines are included. The zero-crossing point marking the approximate boundary between the original disk field and the stellar field slowly expands outward, reaching $\sim 1~\mathrm{au}$ by $t = 4.0~\mathrm{yr}$. 

Although the newly opened field lines within $1~\mathrm{au}$ originate from the stellar magnetosphere rather than the original disk field, they thread the disk vertically and link the dense midplane to the less dense surface layers, just as the original disk field. This connection facilitates the launching of a slower outflow at larger cylindrical radii, which we examine in the following subsection.

\subsection{Disk Wind from Opened Magnetospheric Field}
\label{sec:diskwind}

\begin{figure*}
    \centering
    \includegraphics[width=\textwidth]{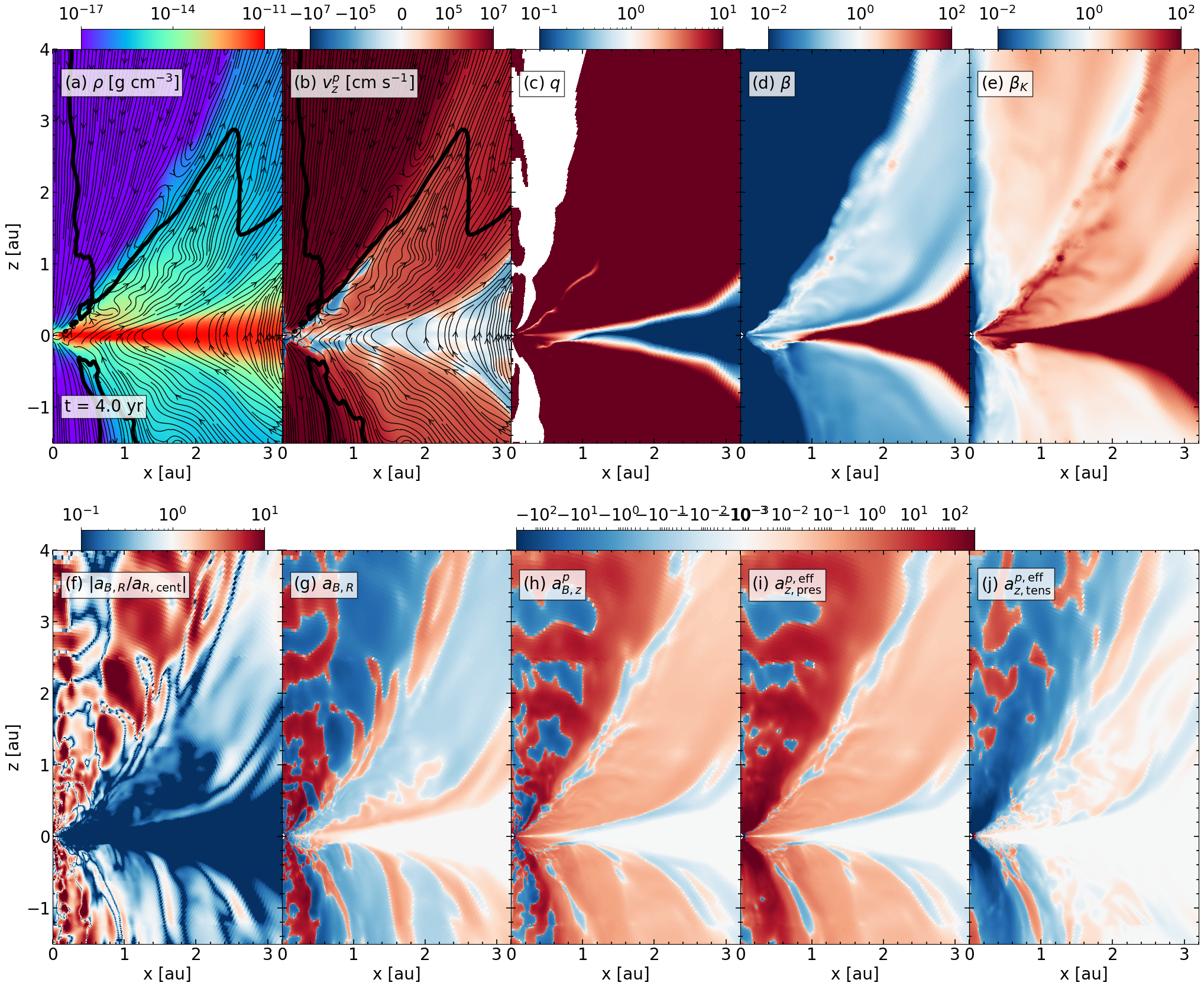}
    \caption{Overview of the outflow, particularly the disk wind at a large cylindrical radius surrounding the jet. {\bf Panel (a) and (b)} show the density and projected $\hat{z}$-direction velocity ($v_z^p$, equ.~\ref{equ:vzp}) respectively, overplotted by the azimuthal averaged fast-magnetosonic surface (the thick solid line); {\bf panel (c)} shows the normalized MRI-turbulence wavelength (equ.~\ref{equ:mri_q}; {\bf panel (d) and (e)} show the plasma-$\beta$ and (equ.~\ref{equ:pbeta}) kinetic-$\beta$ (equ.~\ref{equ:kbeta}) respectively. The lower panels compare different force components in the disk wind. {\bf Panel (f)} shows the ratio between the cylindrical-$\hat{R}$ direction magnetic force and centrifugal force. {\bf Panel (g), (h), (i), and (j)} show the magnitude of the gas acceleration due to magnetic forces. The quantities are total gas acceleration in cylindrical-$\hat{R}$ direction, total magnetic acceleration in $\hat{z}$- direction, the effective magnetic pressure component (equ.~\ref{equ:az_mag_pres_eff}), and the effective magnetic tension component (equ.~\ref{equ:az_mag_tens_eff}) respectively. These disk wind properties and forces show that the disk wind in this model is very similar to the disk wind in \citet{Tu2025b}. } 
    \label{fig:diskwind}
\end{figure*}

The relatively slow, high-density (compared to the polar jet to be described in Section~\ref{sec:magnetospheric_wind_and_jet} below) outflow is driven from a range of disk radii, from $\sim 0.1 - 1$~au, as shown in Figure~\ref{fig:diskwind} where azimuthally averaged flow properties are plotted at a representative time $t=4$~yrs. Since the plotted time is not much longer than the disk orbital period at 1~au,  the outflow driven from the outer part of the radial range is likely still evolving toward a quasi-steady state. With this caveat in mind, we will refer to this outflow as a ``disk wind'' for simplicity hereafter.
Figure~\ref{fig:diskwind} panels (a) and (b) show the density and the projected $\hat{z}$-direction velocity, respectively. 
Any outflow away from the disk midplane has a positive value. The disk wind has a density $\approx 10^{-13}~\mathrm{g\ cm^{-3}}$ near the base (around $r\approx 0.5~\mathrm{au}$), and a density of $\approx 10^{-14}~\mathrm{g\ cm^{-3}}$ at higher altitudes. The outflow velocity of the disk wind is around $10^6~\mathrm{cm\ s^{-1}}$, slower than the jet but still above the escape velocity at high altitudes. The mass loss rate of the disk wind is around a few$\times10^{-8} M_\odot/\mathrm{yr}$ (see Fig.~\ref{fig:acc_theta_3Dstream} below).

\citet{Tu2025b} identify two components for the disk wind: an MRI-active disk wind and a laminar disk wind. In their case, the inner disk wind is MRI-unstable, leading to a turbulent outflow, whereas the outer disk wind is MRI-stable, resulting in a laminar wind. The MRI stability is measured using the dimensionless parameter \citep[][]{Tu2025b}
\begin{equation}
    q\equiv \frac{\lambda}{H},
    \label{equ:mri_q}
\end{equation}
where $H = R/10$ is the disk scale height and $\lambda$ is the most unstable MRI wavelength given by \citep[][]{Zhu2018}
\begin{equation}
    \lambda=2\pi\sqrt{\frac{16}{15}}\frac{v_{A, p}}{\Omega}.
\end{equation}
Here, $v_{A, p}=\sqrt{(B_r^2+B_\theta^2)/4\pi\rho}$ is the poloidal Alfv\'{e}n speed\footnote{Strictly speaking, the Alfv\'{e}n speed should be computed with only the laminar component of the magnetic field. Here, we use the azimuthally-averaged quantities to yield an approximation to the laminar component.} and $\Omega$ is the local orbital frequency. Following \citet{Tu2025b}, we choose the range $0.1 < q < 10$ to denote where MRI is most efficient.

Figure~\ref{fig:diskwind}(c) shows this $q$ parameter, illustrating a zone between the jet and the disk wind in both hemispheres where $q$ is within the optimal range.  The disk wind can thus be divided into the MRI-active disk wind at a small cylindrical radius and the MRI-stable disk wind at a larger cylindrical radius, particularly along the upper hemisphere. However, similar to \citet{Tu2025b}, the MRI-active disk wind eventually merges with the laminar disk wind at high altitudes, where MRI can no longer be sustained. As a result, at these altitudes, the primary distinction is between the jet and the disk wind.

To illustrate the kinematics of the disk wind, we show the plasma-$\beta$ and the kinetic-$\beta$ in Figure~\ref{fig:diskwind}(d) and (e), respectively. They are defined as
\begin{equation}
    \beta = \frac{P}{B^2/8\pi},
    \label{equ:pbeta}
\end{equation}
\begin{equation}
    \beta_K = \frac{\rho v^2/2}{B^2/8\pi},
    \label{equ:kbeta}
\end{equation}
measuring the ratios between the thermal and ram pressures and the magnetic pressure. The plasma-$\beta < 1$ in the disk wind indicates that thermal energy is negligible compared to magnetic energy. 
Plasma-$\beta$ is also comparable to or less than 1 in the inner disk region ($\lesssim$ 1 au), where the rather strong magnetic field drives highly variable, fast (often transonic) accretion streams in the disk and its elevated envelope, as indicated by the blue regions in the $v_z^p$ map (Figure~\ref{fig:diskwind}b), particularly within a radius of $\sim1$ au; the accretion streams can be seen even more clearly in a zoom-in map of radial velocity $v_r$ (not shown, available upon request).
Kinetic-$\beta > 1$ in the disk indicates that kinetic energy (primarily from rotation) dominates the magnetic energy.

The dominance of kinetic energy results in the centrifugal force outweighing the magnetic force in the cylindrical-$\hat{R}$ direction. Figure~\ref{fig:diskwind}(f) and (g) show the ratio between magnetic and centrifugal forces, and the magnitude of the magnetic force in the cylindrical-$\hat{R}$ direction, respectively. The radial-direction magentic force is given by
\begin{equation}
    a_{B, R}=\Big[\frac{1}{4\pi\rho}(\nabla\times\mathbfit{B})\times\mathbfit{B}\Big]_R,
\end{equation}
and the centrifugal force is given by
\begin{equation}
    a_{R, \mathrm{cent}}= \frac{v_\phi^2}{R}.
\end{equation}
While the magnetic force generally points in the $-\hat{R}$ direction (towards the axis), its magnitude is insufficient to counteract the outward centrifugal force. This imbalance leads to an increasing opening angle of the disk wind.

The magnetic force is primarily responsible for driving the outflow in the $\hat{z}$ direction. Figure~\ref{fig:diskwind}(h), (i), and (j) show the projected total magnetic force in the $\hat{z}$ direction, the effective magnetic pressure component, and the effective magnetic tension component, respectively. The projected total magnetic force is given by
\begin{equation}
    a_{B, z}^p = \frac{z}{|z|}\Big[\frac{1}{4\pi\rho}(\nabla\times\mathbfit{B})\times\mathbfit{B}\Big]_z,
    \label{equ:az_mag_force}
\end{equation}
and following \citet{Tu2025b}, we define the projected effective magnetic pressure and tension forces in cylindrical coordinates by removing the $B_z\frac{dB_z}{dz}$ term that does not contribute to overall acceleration:
\begin{equation}
    a_{z, \mathrm{pres}}^{p, \mathrm{eff}} \equiv -\frac{z}{|z|}\frac{1}{4\pi\rho} \Big(\frac{\partial B_\phi}{\partial z}B_\phi + \frac{\partial B_R}{\partial z}B_R\Big),
    \label{equ:az_mag_pres_eff}
\end{equation}
\begin{equation}
    a_\mathrm{z, tens}^{p, \mathrm{eff}} = \frac{z}{|z|}\frac{1}{4\pi\rho}\Big( \frac{\partial B_z}{\partial R}B_R + \frac{\partial B_z}{\partial \phi}B_\phi \Big).
    \label{equ:az_mag_tens_eff}
\end{equation}
It is evident that magnetic pressure drives the disk wind outflow, not the magnetic tension force. This result is consistent with the geometric argument and conclusion in \citet{Tu2025b} that only magnetic pressure can drive an outflow when open magnetic field lines thread a weakly magnetized disk.

\subsection{Opened Stellar Field}
\label{sec:openstarfield}

As discussed in Sec.~\ref{sec:mag_flux_evol}, part of the stellar magnetospheric field opens up. Half of the opened field threads the disk and becomes a ``disk field''; the other half remains attached to the star and becomes an opened stellar field. 

\begin{figure*}
    \centering
    \includegraphics[width=\textwidth]{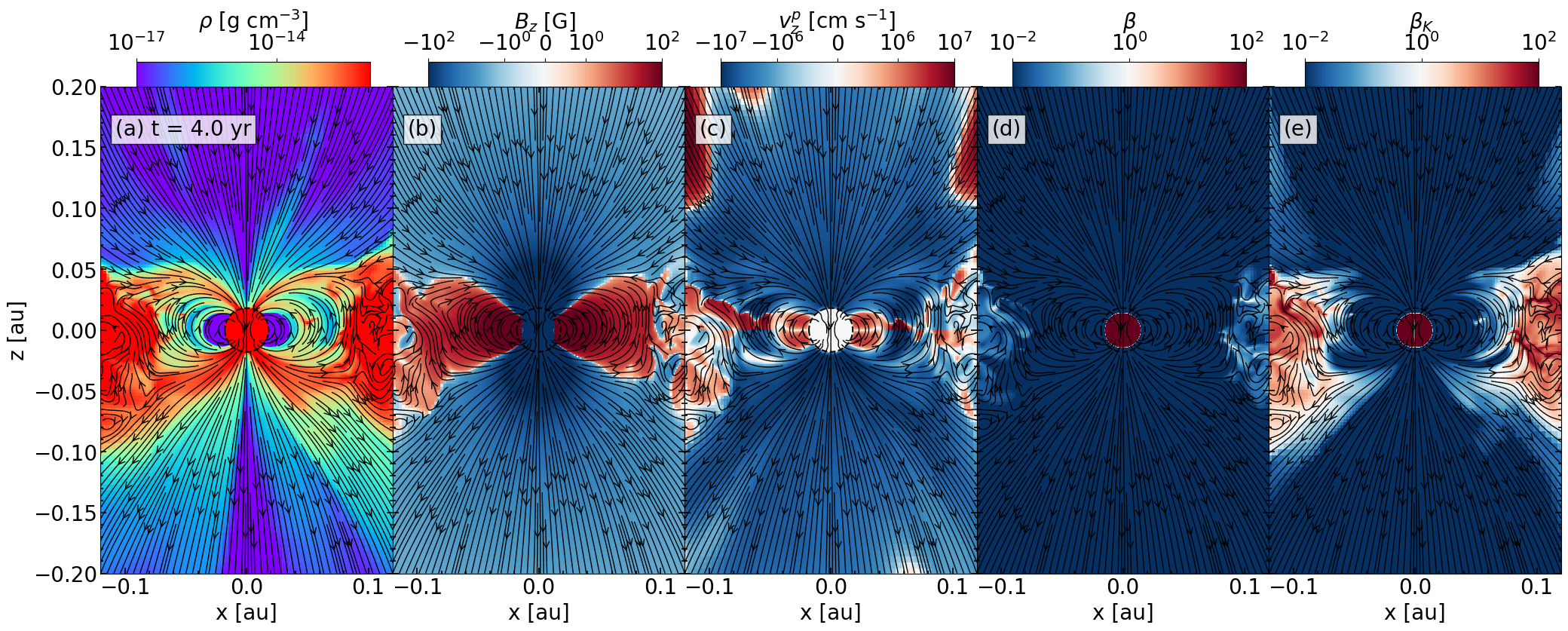}
    \caption{A zoom-in view focusing on the permanently closed stellar magnetosphere (around the midplane) and the opened stellar magnetic field threading the star (in the polar region above and below the star). {\bf Panel (a)} shows the real density traced by the ``passive scalars'', highlighting the magnetospheric accretion streams. The sphere in the middle is the inner boundary $r_\mathrm{fix}$, where a star resides  (see sec.~\ref{sec:method} for a more detailed description of the setup). {\bf Panels (b) and (c)} show the $B_z$ field strength and $v_z^p$ (equ.~\ref{equ:vzp}) respectively, illustrating the magnetic and dynamical properties of the region close to the star. {\bf Panels (d) and (e)} show the plasma-$\beta$ (equ.~\ref{equ:pbeta}) and kinetic-$\beta$ (equ.~\ref{equ:kbeta}) respectively, indicating the magnetosphere and the polar cavity are both magnetically dominated.} 
    \label{fig:starfield}
\end{figure*}

To illustrate the geometry of the opened stellar magnetic field, Figure~\ref{fig:starfield}(a) presents a density slice through the meridional plane, overlaid with magnetic field lines. 
On the displayed scale, the magnetic field in the polar regions above and below the star points in the $-\hat{z}$ direction, opposite to the field threading the disk (e.g., Figure~\ref{fig:diskwind}), as expected. 

The properties of this downward magnetic field and the gas kinematics are shown in Figure~\ref{fig:starfield}(b) and (c), respectively. The magnetic field strength is particularly strong in the polar cavity, reaching approximately 20 Gauss at $z = \pm0.1~\mathrm{au}$. The gas in the polar cavity immediately above and below the stellar magnetosphere rapidly falls toward the star because of strong gravity and insufficient rotational support. Given the low density and strong magnetic field in this region, it is unsurprising that the polar cavity is entirely magnetically dominated, with plasma-$\beta$ and kinetic-$\beta$ values much lower than 1 (Figure~\ref{fig:starfield}[d] and [e]).

Although the majority of the open stellar field does not directly contribute to jet launching, it serves as a structural backbone for the jet in the magnetically dominated region. As we demonstrate in Section~\ref{sec:magnetospheric_wind_and_jet}, the jet is primarily driven by toroidal magnetic pressure. However, a magnetic tower consisting predominantly of toroidal loops is inherently susceptible to kink instabilities \citep{Moll2008, Huarte-Espinosa2012}. The strong open stellar magnetic field in the polar region, which accounts for most of the open stellar flux, helps stabilize the jet by threading these loops along the polar axis, mitigating instabilities.

\subsection{Magnetospheric jet}
\label{sec:magnetospheric_wind_and_jet}

\begin{figure*}
    \centering
    \includegraphics[width=\textwidth]{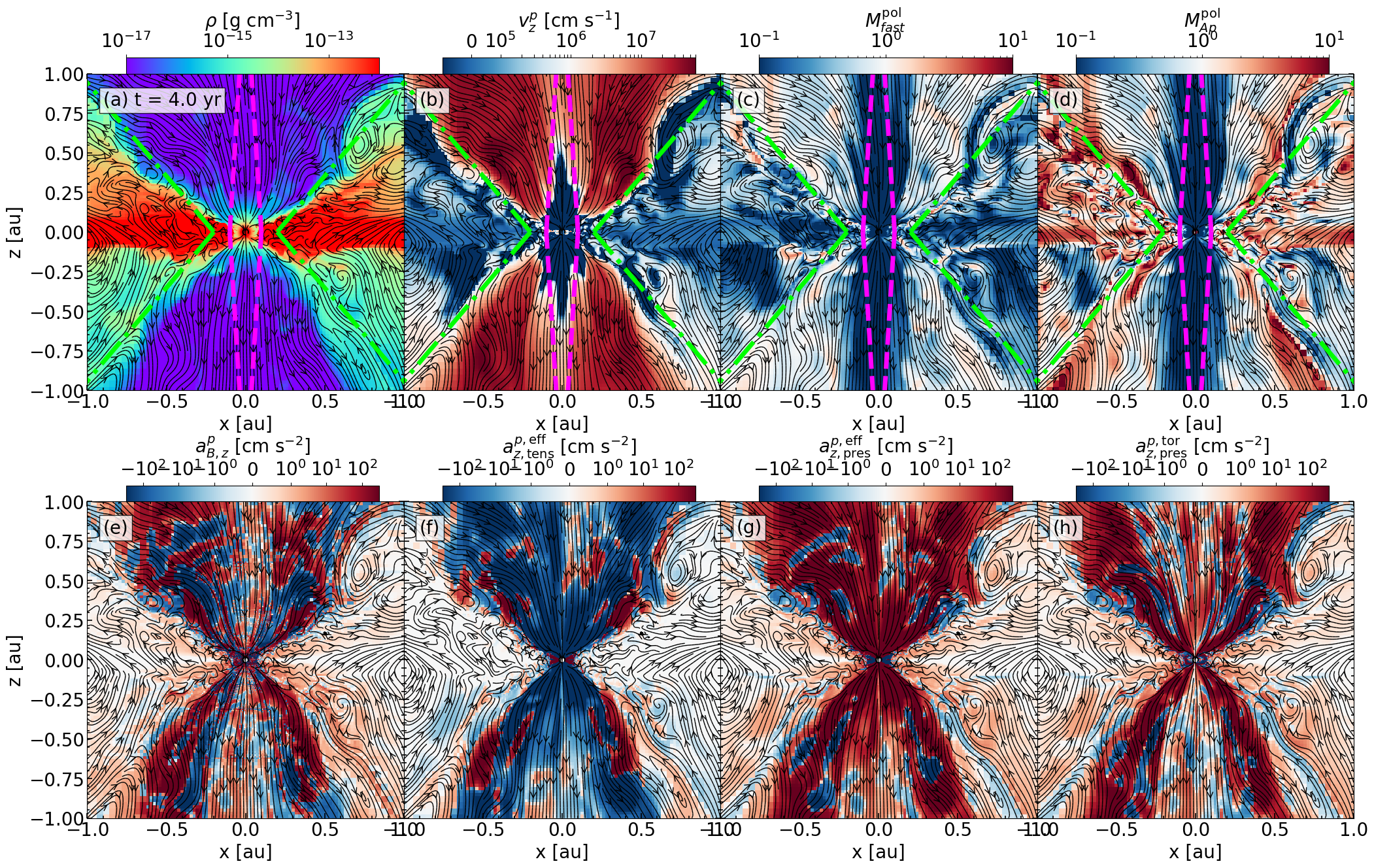}
    \caption{Overview of the jet. Upper panels focus on the magneto-hydrodynamic properties of the jet; lower panels focus on the magnetic forces. \textbf{Panels (a) and (b)} show the density and projected $\hat{z}$-direction ($v_z^p$, equ.~\ref{equ:vzp}) respectively, showing the jet lies in the low-density polar cavity. 
    \textbf{Panels (c) and (d)} show the poloidal fast Mach number and poloidal Alfv\'en Mach number, respectively, showing the jet is a magnetically-dominated structure.
    \textbf{Panels (e), (f), (g), and (h)} show the total magnetic force, the effective magnetic tension force (equ.~\ref{equ:az_mag_tens_eff}), the effective magnetic pressure force (equ.~\ref{equ:az_mag_pres_eff}), and the toroidal magnetic pressure force (equ.~\ref{equ:az_mag_pres_tor}) respectively.}
    \label{fig:magnetospheric_jet_props}
\end{figure*}

The jet is launched from the interaction between the star, its magnetosphere, and the surrounding disk, and propagates in the low-density polar cavity. Figure~\ref{fig:magnetospheric_jet_props} outlines the jet’s location and key characteristics. At lower altitudes ($\lesssim 0.5~\mathrm{au}$), the jet is confined between an infall region along the opened stellar field near the polar axis (approximately delineated by the magenta line) and the disk wind/atmosphere at larger cylindrical radii (indicated by the green dashed line). At higher altitudes, the jet expands to fill the polar cavity, as shown in Figure~\ref{fig:diskwind}[b]. The mass loss rate of the jet remains around $2\times10^{-9} M_\odot/\mathrm{yr}$ in our model (see Fig.~\ref{fig:acc_theta_3Dstream} below).

To assess the dynamic importance of the magnetic field, we show in Figure~\ref{fig:magnetospheric_jet_props}(c) and (d) the poloidal fast magnetosonic Mach number ($M_\mathrm{fast}^\mathrm{pol} = v_\mathrm{pol} / v_\mathrm{fast}$) and the poloidal Alfv\'en Mach number ($M_\mathrm{Ap}^\mathrm{pol} = v_\mathrm{pol} / v_\mathrm{A, pol}$), respectively. The poloidal Alfv\'en speed $v_\mathrm{A, pol}$ is computed using only the poloidal component of the magnetic field. Both $M_\mathrm{fast}^\mathrm{pol}$ and $M_\mathrm{Ap}^\mathrm{pol}$ are less than 1 in the jet-launching polar region, indicating the jet launching process is magnetically-dominated, as expected. However, we note that the jet becomes kinematically dominated at large heights (Figure~\ref{fig:diskwind}[e]) as energy is transferred from the magnetic field to the gas (see Section~\ref{sec:conversion} below for more details). 
The jet is well collimated at large distances (see Fig.~\ref{fig:3D} and Fig.~\ref{fig:StellarWind}e), likely primarily by lateral confinement from the denser, more slowly moving disk wind
surrounding it. Internal ``hoop" stresses from its toroidal magnetic field may also contribute (see Fig.~\ref{fig:diskwind}g, which shows that the magnetic force in the cylindrical radial direction points inward toward the axis within the jet).

Since both $M_\mathrm{fast}^\mathrm{pol}$ and $M_\mathrm{Ap}^\mathrm{pol}$ are below order unity in the jet-launching polar cavity, we focus on the magnetic forces to establish a foundation for our detailed examination of the jet-launching mechanism in section ~\ref{sec:jet-launching-mechanism}. The lower panels of Figure~\ref{fig:magnetospheric_jet_props} illustrate the magnetic forces in the jet region. Figure~\ref{fig:magnetospheric_jet_props}[e] shows the total magnetic force, revealing that the acceleration of the outflow occurs primarily along the surface of the vertically extended disk, which also marks the boundary of the jet-launching cavity. 
Above approximately 1 au in both hemispheres, acceleration becomes less coherent (with alternating regions of acceleration and deceleration\footnote{These accelerations and decelerations can potentially introduce shocks in the jet, which can result in heating of the jet.}), indicating that the bulk of the flow acceleration has largely completed beyond this altitude. This is confirmed by velocities in the $\hat{z}$ direction reaching several $10^7~\mathrm{cm~s^{-1}}$ below 1 au, with minimal subsequent increase in velocity above this height.

As demonstrated in \citet{Tu2025b}, in systems with ordered, large-scale magnetic fields threading a weakly magnetized disk, only magnetic pressure forces, not magnetic tension forces, can launch outflows in global models. Our simulations confirm this finding. Figure~\ref{fig:magnetospheric_jet_props}(f) and (g) display the effective magnetic tension acceleration (equ.~\ref{equ:az_mag_tens_eff}) and effective magnetic pressure acceleration (equ.~\ref{equ:az_mag_pres_eff}), respectively. Magnetic tension acceleration generally points toward the midplane, while magnetic pressure acceleration points away from it. More specifically, Figure~\ref{fig:magnetospheric_jet_props}(h) shows the acceleration due to toroidal magnetic pressure gradient, defined as

\begin{equation}
    a^{p, \mathrm{tor}}_{z, \mathrm{pres}} = -\frac{z}{|z|}\frac{1}{4\pi\rho}\frac{dB_\phi}{dz}B_\phi.
    \label{equ:az_mag_pres_tor}
\end{equation}

The toroidal magnetic pressure accounts for most of the magnetic pressure-driven jet acceleration. In the following section, we will examine the mechanism generating this crucial toroidal magnetic field (Section~\ref{sec:jet-launching-mechanism}).

\section{Jet launching mechanism: ``Load-fire-reload''}
\label{sec:jet-launching-mechanism}

This section presents a detailed description and analysis of the jet-launching mechanism in our model. The jet is launched by magnetic field lines connecting the star and the disk (we name them ``two-legged'' field lines, see Figure~\ref{fig:init_open}[b] for an illustration), through a cyclic process comprising three distinct phases: load, fire, and reload. Each phase corresponds to characteristic changes in the magnetic field configuration and dynamics. We emphasize here that each cycle occurs on extremely short time scales, and only their global effects can be captured. In the following subsections, we describe each step in detail and analyze how they collectively drive the jet-launching behavior observed in the simulation.

\subsection{Load: charging $B_\phi$ by star-disk differential rotation}
\label{sec:load}

\begin{figure*}
    \centering
    \includegraphics[width=\linewidth]{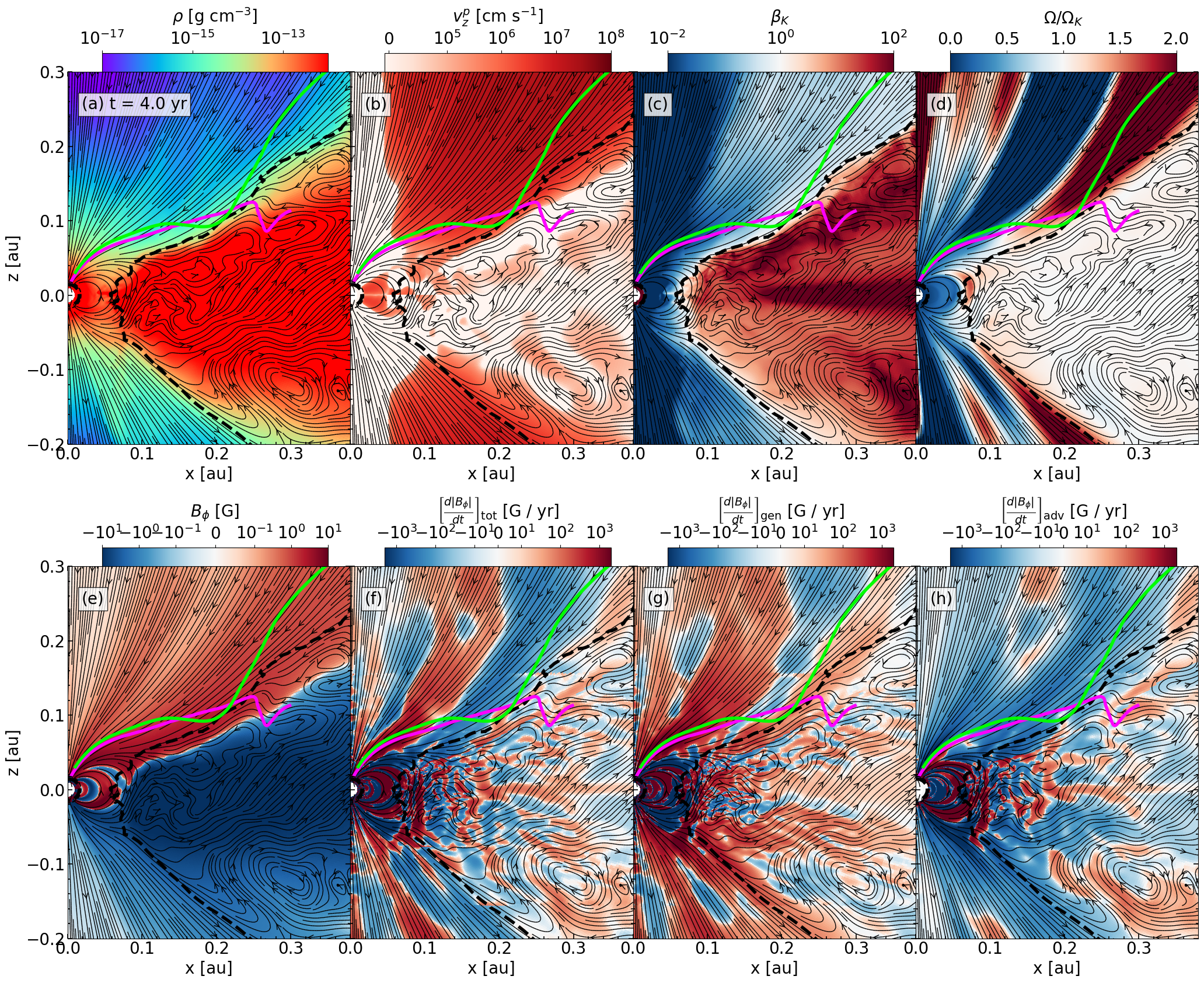}
    \caption{Toroidal magnetic field generation in the jet-launching region. The quantities plotted are azimuthally averaged. \textbf{Panel (a) and (b)} show the density and projected $\hat{z}$-direction velocity ($v_z^p$, equ.~\ref{equ:vzp}) respectively, showing the density structure and the location of the jet. \textbf{Panel (c)} shows the kinetic-$\beta$ (equ.~\ref{equ:kbeta}), highlighting the transition from the magnetically-dominated polar cavity to the mass-dominated disk. \textbf{Panel (d)} shows the angular velocity, normalized by the local Keplerian value. The lower panels show toroidal magnetic field dynamics. \textbf{Panel (e)} shows the toroidal magnetic field; \textbf{Panels (f), (g), and (h)} show the total change in toroidal field strength (equ.~\ref{equ: change Bphi strength}), toroidal field generation by differential rotation (equ.~\ref{equ:Bphi generation}), and toroidal field advection by flux-freezing (equ.~\ref{equ:Bphi advection}) respectively. The black dashed contour highlights the $\beta_K = 1$ surface. The magenta and green field lines are 2D projections of representative 3D field lines shown in Figure~\ref {fig:jet_twist_fieldlines}. An animated version is available at 
    \url{https://figshare.com/s/e90cdd7f3a49d07ffbc1}. The movie is 25 seconds long, showing the gas properties on the upper panels, and the toroidal field dynamics on the lower panels.}
    \label{fig:jet_twist}
\end{figure*}

To illustrate the jet-launching mechanism, we zoom in on the inner part of the simulation in Figure~\ref{fig:jet_twist}, which shows the azimuthally averaged gas and magnetic field properties. Panels (a), (b), and (c) show the density, projected $\hat{z}$-direction velocity $v_z^p$, and kinetic-$\beta$, respectively. The jet resides in the low-density polar cavity, where $\rho\lesssim10^{-14}~\mathrm{g\ cm^{-3}}$, $v_z^p > 10^7~\mathrm{cm\ s^{-1}}$ and $\beta_K < 1$. The low $\beta_K$ highlights the magnetic dominance in the jet-launching polar cavity. A black dashed contour traces $\beta_K = 1$, the rough boundary between the high-density disk wind/atmosphere and the low-density jet.

On the scale shown in Figure~\ref{fig:magnetospheric_jet_props}, the jet is driven by toroidal magnetic pressure primarily along the polar cavity–disk boundary. These toroidal fields arise where differential rotation occurs along a field line, most effectively on lines threading both the star, where the magnetically dominated segment co-rotates with the stellar surface, and the disk, where the gas forces the field to move at the local Keplerian velocity (the ``two-legged'' field lines). These field lines connecting the star and dense disk generate toroidal fields through angular velocity gradients along their length, as illustrated in Figure~\ref{fig:jet_twist}(d). Across the $\beta_K = 1$ boundary, the angular velocity shifts from nearly Keplerian in the disk to super-Keplerian above it, highlighting the shear that twists field lines and generates a toroidal magnetic field.

To showcase a ``two-legged'' magnetic field participating in this process, Figure~\ref{fig:jet_twist} presents a 2D projection of a representative field line (solid magenta). Anchored on the star, the line transfers angular momentum to polar gas in the low-$\beta$ region, forcing super-Keplerian rotation. Its other end remains embedded in the nearly Keplerian disk. The resulting differential rotation rapidly produces a strong azimuthal field [Fig.~\ref{fig:jet_twist}(e)].

\begin{figure}
    \centering
    \includegraphics[width=\linewidth]{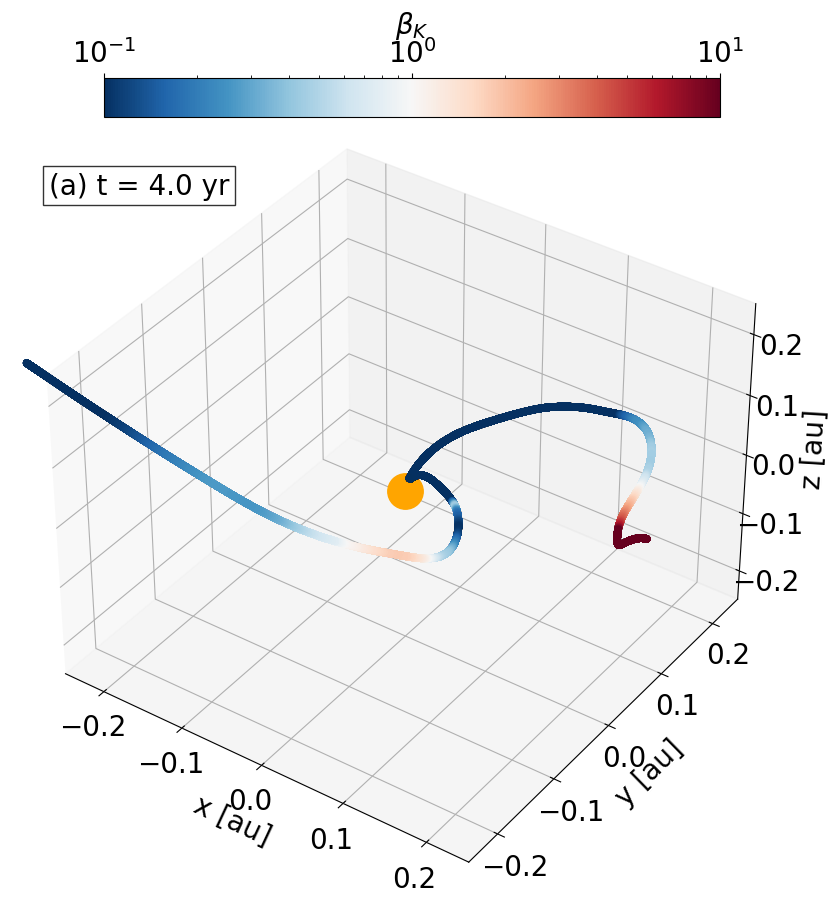}
    \caption{3D rendition of the two fieldlines highlighted in Figure~\ref{fig:jet_twist}, colored by the local kinetic-$\beta$ (equ.~\ref{equ:kbeta}). The yellow sphere at the center represents the star. The magenta fieldline in Figure~\ref {fig:jet_twist} is located on the right, with one low kinetic-$\beta$ end on the star and one high kinetic-$\beta$ end in the disk. The green field line from Figure\ref{fig:jet_twist} is shown on the left: most of this field line resides in regions where $\beta_K < 1$, indicating magnetic dominance, except for a localized segment with $\beta_K > 1$, where kinetic energy becomes dominant.}
    \label{fig:jet_twist_fieldlines}
\end{figure}

\begin{figure*}
    \centering
    \includegraphics[width=\linewidth]{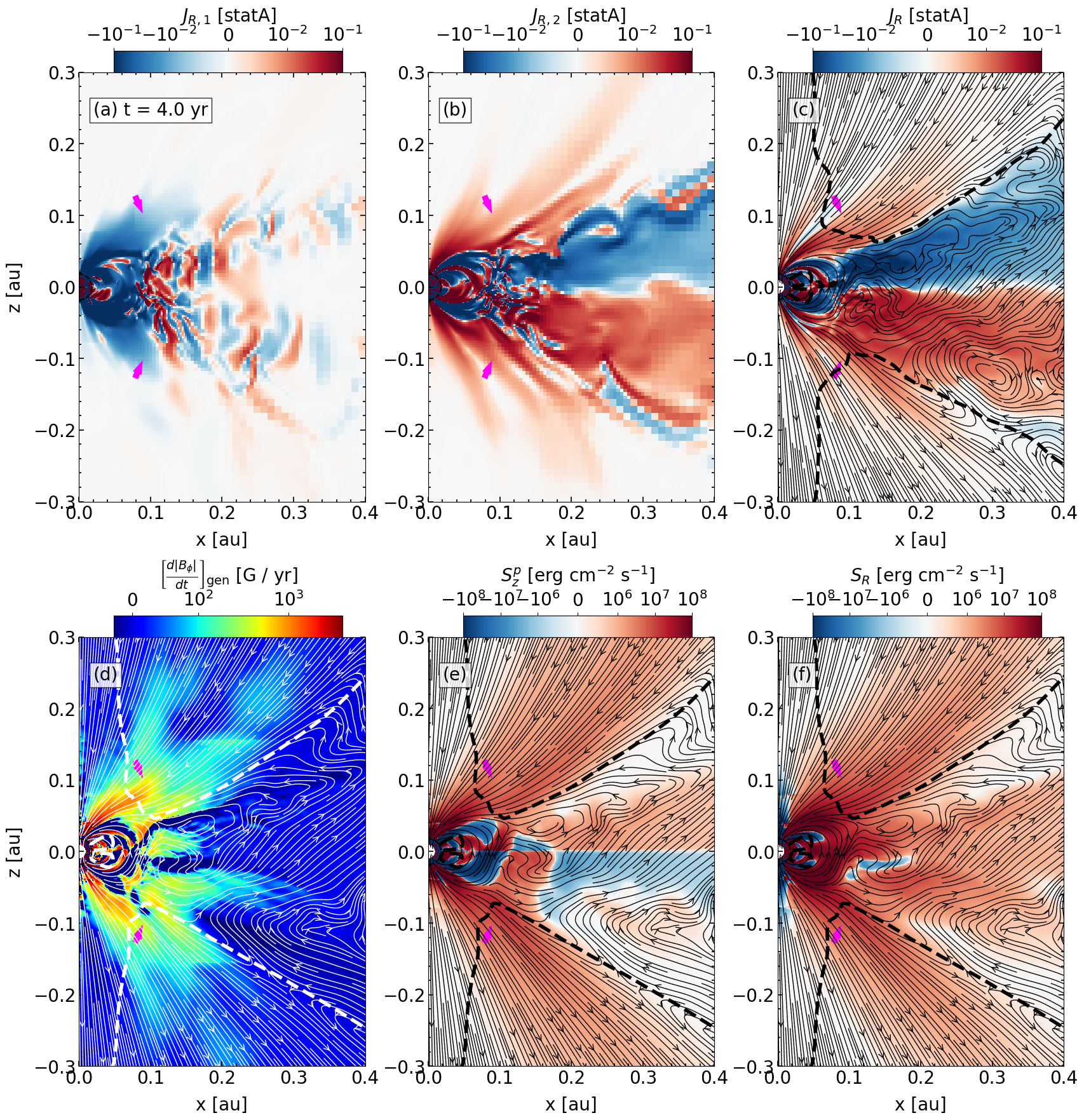}
    \caption{Magnetic energy evolution in the jet-launching region. The top panels illustrate the cylindrical $\hat{R}$-direction current and its components. Given the toroidal magnetic field structure (Figure~\ref{fig:jet_twist}[e]), a positive current is required to exert a force directly away from the disk. \textbf{Panels (a) and (b)} show meridional slices of the two components of the $\hat{R}$-directed current (equ.~\ref{equ:JR1} and \ref{equ:JR2}), respectively. The component due to azimuthal asymmetry (equ.~\ref{equ:JR1}) points toward the midplane, while the component arising from global field geometry (equ.~\ref{equ:JR2}) points away from the disk. \textbf{Panel (c)} shows the azimuthally averaged current at a representative time, revealing a positive current along the surfaces of both the magnetosphere and disk—sufficient to drive outflows. To reduce transient effects and emphasize large-scale features, panels (d), (e), and (f) present time-averaged (over $\pm$0.1 yr centered on 4.0 yr) and azimuthally averaged quantities. \textbf{Panel (d)} shows toroidal field generation concentrated along narrow streams at the magnetosphere and disk surfaces, highlighting the origin of the current. \textbf{Panels (e) and (f)} display the projected Poynting flux in the $\hat{z}$- and $\hat{R}$-directions, respectively, demonstrating how magnetic energy generated near the star is transported outward to power the jet. The dashed contours in panels (c)–(f) outline the approximate extent of the jet.}
    \label{fig:current_sheet}
\end{figure*}

The strong azimuthal magnetic field empowers a jet, accelerating the gas in the vertical $\hat{z}$-direction. Consequently, this azimuthal magnetic field is constantly advected away by gas motion. Quantitatively, magnetic field generation and removal in this nearly ideal MHD region (due to ionization by high temperature) are governed by the induction equation (equ.~\ref{equ:mhd induction}). In a cylindrical coordinate, the change in azimuthal magnetic field ($\big[\frac{\partial B_\phi}{\partial t}\big]_\mathrm{tot}$) can be divided into two processes: generation of $B_\phi$ from $B_z$ and $B_R$ by differential rotation,
\begin{equation}
    \Big[\frac{\partial B_\phi}{\partial t}\Big]_\mathrm{gen} = 
    \frac{\partial (v_\phi B_z)}{\partial z} 
    + \frac{\partial (v_\phi B_R)}{\partial R},
    \label{equ:Bphi generation}
\end{equation}
and advection of $B_\phi$ by gas motion,
\begin{equation}
    \Big[\frac{\partial B_\phi}{\partial t}\Big]_\mathrm{adv} = 
    - \frac{\partial (v_z B_\phi)}{\partial z} 
    - \frac{\partial (v_R B_\phi)}{\partial R}.
    \label{equ:Bphi advection}
\end{equation}
Since the toroidal magnetic field strength (and the magnetic pressure associated with it) is what ultimately powers jet-launching, we focus on the change in toroidal magnetic field strength by defining
\begin{equation}
    \frac{\partial |B_\phi|}{\partial t} \equiv \frac{B_\phi}{|B_\phi|}\frac{\partial B_\phi}{\partial t},
    \label{equ: change Bphi strength}
\end{equation}
where a positive value denotes an increase in the absolute value of the toroidal magnetic field strength and a negative value denotes a decrease. Figure~\ref{fig:jet_twist}(f), (g), and (h) shows $\big[\frac{\partial |B_\phi|}{\partial t}\big]_\mathrm{tot}$, $\big[\frac{\partial |B_\phi|}{\partial t}\big]_\mathrm{gen}$, and $\big[\frac{\partial |B_\phi|}{\partial t}\big]_\mathrm{adv}$ respectively\footnote{ Note that the flow quantities shown in Fig.~\ref{fig:jet_twist} are first computed in each cell and then averaged azimuthally. Non-azimuthally averaged quantities yield qualitatively similar results but are much noisier.}. Although $|B_\phi|$ decreases in magnitude in a large part of the jet-launching region at the time shown, $|B_\phi|$ switches rapidly between net gain and loss of strength over time, resulting in a quasi-steady state. The reader is encouraged to view the animated version of Figure~\ref{fig:jet_twist} to see this rapid change between gaining and losing $|B_\phi|$.

The rapid change in the net change in $\big[\frac{\partial |B_\phi|}{\partial t}\big]_\mathrm{tot}$ does not translate to a rapid change in the $\big[\frac{\partial |B_\phi|}{\partial t}\big]_\mathrm{gen}$ nor $\big[\frac{\partial |B_\phi|}{\partial t}\big]_\mathrm{adv}$. Generation through differential rotation ($\big[\frac{\partial |B_\phi|}{\partial t}\big]_\mathrm{gen}$) stays mostly positive on the magnetically dominated side of the $\beta_K = 1$ boundary, and advection through gas motion ($\big[\frac{\partial |B_\phi|}{\partial t}\big]_\mathrm{adv}$) stays mostly negative, illustrating the required toroidal magnetic field for jet-launching is generated in situ by differential rotation along magnetic field lines. Quantitatively, the magnitude of $\big[\tfrac{\partial |B_\phi|}{\partial t}\big]_\mathrm{gen}$ reaches $\sim 10^3~\mathrm{G~yr^{-1}}$ along the jet-launching disk surface, corresponding to a \emph{toroidal} field generation time scale $<10^{-2}~\mathrm{year}$. In comparison, the stellar surface \emph{poloidal} dipole magnetic field strength is of the order of $10^3~\mathrm{G}$, suggesting that the \emph{toroidal} field generation and advection proceed on remarkably short timescales.

The $\big[\frac{\partial |B_\phi|}{\partial t}\big]_\mathrm{gen}$ term being predominantly positive in the jet-launching region indicates that toroidal field generation is not limited to field lines like the magenta fieldline highlighted in Figure~\ref{fig:jet_twist}, where one footpoint is anchored in the disk and the other on the star. In fact, any differential rotation between the dense disk gas and the stellar magnetosphere can generate a toroidal magnetic field capable of launching an outflow, even if the field line is not fully embedded in the disk. An example of this ``touch-and-go'' two-legged field lines is shown in Figure~\ref{fig:jet_twist_fieldlines}, where a segment along an otherwise open field line becomes kinetically dominated ($\beta_K > 1$) as it approaches the disk. This ``touch-and-go'' interaction is facilitated by the elevated disk atmosphere, which allows the field line to interact with the disk surface without the need to reach close to the disk midplane. The local interaction between the disk and the stellar field generates differential rotation, which in turn produces the toroidal field responsible for jet launching.

Similar field lines threading both the disk and the star to the one highlighted in Figure~\ref{fig:jet_twist}(d) can be found in the 3D plot (and the associated animation) of the field lines in the magnetosphere and surrounding regions (see Figure~\ref{fig:3D_magnetosphere} below) along many azimuthal directions, allowing the jet to be driven continuously in the simulation. 


\subsection{Fire: releasing magnetic energy by global field geometry}
\label{sec:fire}

Another critical consequence of star-disk differential rotation is the emergence of strong current sheets, particularly in the cylindrical $\hat{R}$ direction, which are essential for converting magnetic energy stored in the toroidal field into jet-driving forces. Specifically, the abrupt change in magnetic field geometry by the differential rotation between the star and the disk leads to the generation of a significant radial current density, $J_R$, expressed as:
\begin{equation}
    J_R = \frac{c}{4\pi}\Big(\frac{\partial B_z}{\partial\phi} - \frac{\partial B_\phi}{\partial z}\Big).
    \label{equ:JcylR}
\end{equation}
This current has two physically distinct contributions:
\begin{equation}
    J_{R, 1} = \frac{c}{4\pi}\frac{\partial B_z}{\partial\phi},
    \label{equ:JR1}
\end{equation}
arises from azimuthal asymmetries in $B_z$, while the second term,
\begin{equation}
    J_{R, 2} = - \frac{c}{4\pi}\frac{\partial B_\phi}{\partial z},
    \label{equ:JR2}
\end{equation}
stems from vertical gradients in the toroidal magnetic field. Since azimuthal averaging eliminates $J_{R,1}$ by construction, we examine both terms in meridional slices (see Figure~\ref{fig:current_sheet}[a,b]) to evaluate their roles in driving outflows.

For a magnetic force to support jet launching in both hemispheres, the condition $\frac{z}{|z|} J_R B_\phi > 0$ must be satisfied, such that the Lorentz force points away from the midplane.  This is generally satisfied when $J_R > 0$, given the sign of $B_\phi$ in Figure~\ref{fig:jet_twist}[e], which is positive in the polar cavity of the upper hemisphere and negative in that of the lower hemisphere in jet-launching inner disk region. While $J_{R,1}$ tends to oppose the jet by introducing a restoring force toward the midplane, $J_{R,2}$ contributes positively to $J_R$, overcoming the restoring force by $J_{R, 1}$, and aligning with the outflow direction. 

This jet-driving current is generated locally along the surface of the closed magnetosphere and the disk as a consequence of the rapid changes in the toroidal magnetic field geometry in the vertical direction. As discussed in Sec.~\ref{sec:load}, the toroidal field is most rapidly generated when the field line has one end rooted in the star and the other embedded in the disk. A comparatively strong toroidal field is produced in this narrow layer where such ``two-legged" field lines reside. Above this region, in the low-density polar cavity, the strong magnetic field keeps the fieldlines roughly poloidal, and the toroidal component remains weak. Below this region is the mass-dominated disk, where the magnetic effect is more diminished, providing a solid foundation for the outflow. 

As a result, the field lines within this intermediate layer, subject to significant differential rotation, are continually over-twisted relative to those above and below. This localized over-twisting establishes steep vertical gradients in the toroidal magnetic field, giving rise to a strong radial current sheet ($J_R$) essential for jet launching. Because this differential twisting is inherently highly variable in space and time, we illustrate it in Figure~\ref{fig:current_sheet}(d) using the azimuthally and temporally averaged quantity $\left[\frac{\partial B_\phi}{\partial t}\right]_\mathrm{gen}$. The colormap highlights the extremely rapid generation rate of the toroidal magnetic field ($\gtrsim 10^3~\mathrm{G~yr^{-1}}$) in the layer above the closed magnetosphere and disk atmosphere, with more positive values indicating faster local toroidal field generation. The peaks in the toroidal field generation rate, pointed by arrows in Figure~\ref{fig:current_sheet}(d), align with the location of the $J_R$ current sheet at the disk surface. However, as shown in  Figure~\ref{fig:current_sheet}(d), the gas in the most rapid toroidal field generating region near the star is infalling, rather than outflowing (the outflow region is outlined in white in Figure~\ref{fig:current_sheet}[d]). This raises a natural question: how does the energy generated in this inner region reach the outflow zone to drive the jet? 

The answer lies in the Poynting flux, defined as
\begin{equation}
    \mathbfit{S}=-\frac{1}{4\pi}(\mathbfit{v}\times \mathbfit{B})\times \mathbfit{B},
\end{equation}
where $\mathbfit{v}$ and $\mathbfit{B}$ are the gas velocity and magnetic field respectively. The Poynting flux enables magnetic energy to travel faster than the gas itself. Even in regions of infall, magnetic fields can transport energy outward, effectively ``advecting'' it upstream into the jet-launching region. To illustrate this process, we show in Figure~\ref{fig:current_sheet}(d, e) the Poynting flux in $\hat{z}$ direction ($S_z^p=z/|z|~\mathbfit{S}\hat{\mathbfit{z}}$, projected in the same way as the $\hat{z}$ direction velocity $v_z^p$, equ.~\ref{equ:vzp}) and cylindrical-$\hat{R}$ directions ($S_R=\mathbfit{S}\hat{\mathbfit{R}}$), respectively. Even in the infall region around the polar axis, magnetic energy can be transported away from the star's vicinity to the outflow region.

In summary, the combined effect of the $J_R$ current sheet (Figure~\ref{fig:current_sheet}[c]) and the amplified toroidal field (Figure~\ref{fig:jet_twist}[e]) generates a vertical ($\hat{z}$-direction) magnetic force, accelerating gas to high poloidal velocities. 

\begin{figure*}
    \centering
    \includegraphics[width=\linewidth]{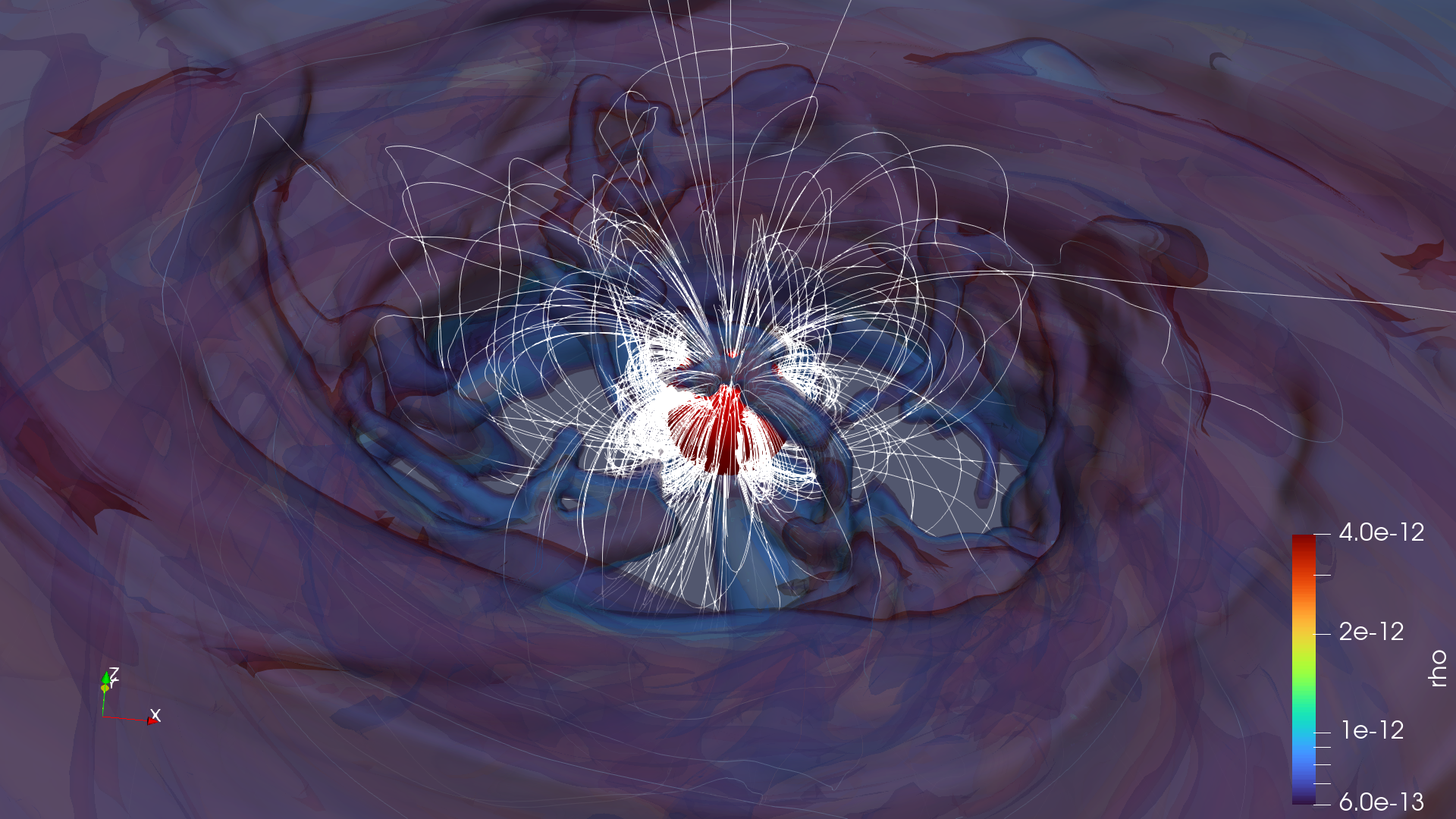}
    \caption{3D rendition of the protostar, accreting magnetosphere, and magnetic field lines. The disk is shown as semi-transparent density isosurfaces at around $10^{-12}~\mathrm{g~cm^{-3}}$. The protostar is represented by the central red solid sphere. White lines depict magnetic field lines, including permanently closed magnetospheric loops and twisted field lines resulting from magnetosphere–disk interactions (e.g., the prominent twisted fieldline near the top-center of the image). An animated version of this figure is available at: 
    \url{https://figshare.com/s/413012328cf225f86a67}. The movie is 27 seconds long, showing the magnetic field lines near the magnetosphere.}
    \label{fig:3D_magnetosphere}
\end{figure*}

\subsection{Reload: replenishing jet gas by magnetic reconnection}
\label{sec:reload}

The ``two-legged'' magnetic field lines, such as those shown in Figure~\ref{fig:jet_twist}(e), stretch and may open during outflow launching. To sustain a continuous jet (Figure~\ref{fig:3D}), these field lines must reconnect and revert to a ``two-legged'' configuration, ``reloading'' with mass by magnetic reconnection between oppositely directed field lines. As discussed in Sections~\ref{sec:mag_flux_evol} and~\ref{sec:openstarfield}, the global magnetic topology at late times when a persistent outflow is launched consists of downward-pointing original magnetospheric field lines that are opened but remain connected to the star (referred to as ``the opened stellar field lines" for short) and upward-pointing original magnetospheric field lines that are opened but thread through the (inner) disk (but not the star). These oppositely directed open field lines are expected to preferentially reconnect at their interface. Although individual reconnection events are difficult to isolate in a highly variable and rapidly evolving 3D system, we present qualitative evidence below to confirm reconnection along this boundary.  

\begin{figure*}
    \centering
    \includegraphics[width=\linewidth]{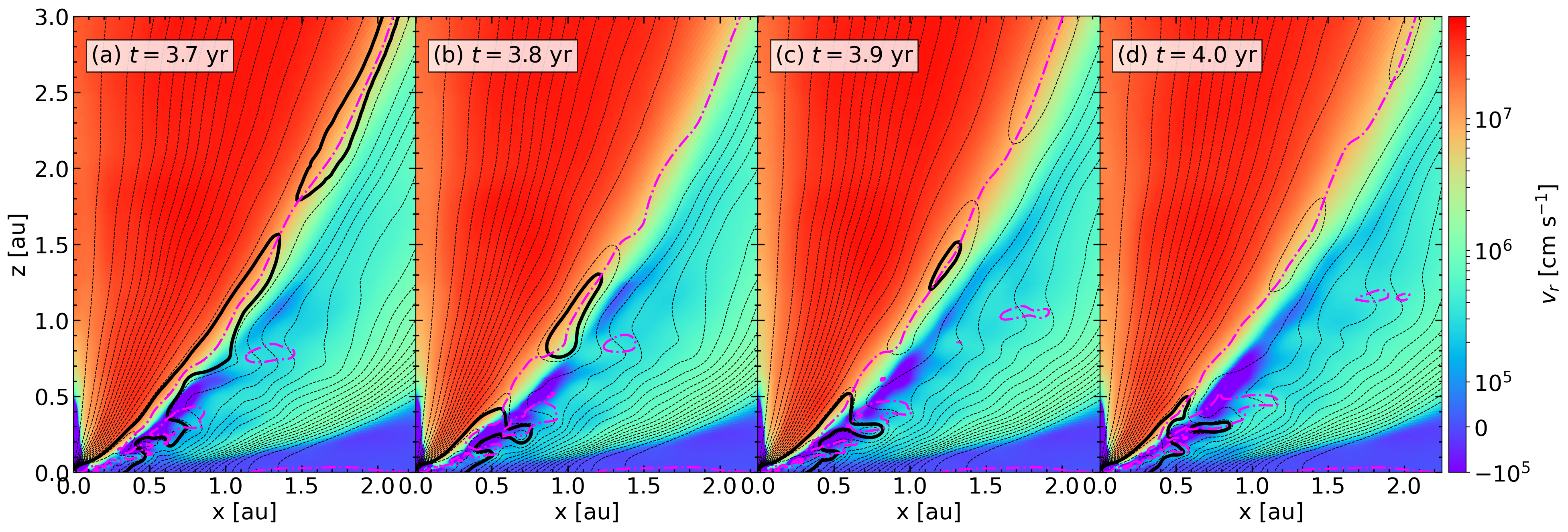}
    \caption{Illustration of the ``reload'' stage where magnetic reconnection resets the field geometry. While individual reconnection event occurs rapidly and is unrealistic to show explicitly in 3D simulations, we illustrate the cumulative effect using azimuthally integrated magnetic flux surfaces. The panels present a time sequence of the flux surface evolution. The background color shows the radial velocity in spherical coordinates ($v_r$), highlighting the jet's location. Black lines represent azimuthally averaged magnetic flux surfaces, which can also be interpreted as approximate magnetic field line locations. The thick black line shows magnetic reconnection by tracing a representative field line along the $B_r = 0$ surface, which is marked by the magenta dash-dot line.} 
    \label{fig:Br_flux}
\end{figure*}

Figure~\ref{fig:Br_flux}(a)-(d) presents the time evolution of the azimuthally averaged radial velocity $v_r$ at $t = 3.7$, $3.8$, $3.9$, and $4.0~\mathrm{yr}$. To track the evolution of the magnetic field, we trace field lines using contours of the radial magnetic flux:  
\begin{equation}
    \Phi_{B, r}(r, \theta) = \int_0^\theta \int_0^{2\pi} B_r r^2 \sin\theta' \, d\phi d\theta',  
\end{equation}
where constant $\Phi_{B, r}$ contours approximate field line locations, as indicated by the black dashed curves in Figure~\ref{fig:Br_flux}. These contours represent azimuthally averaged field configurations, capturing only the global magnetic evolution. In reality, three-dimensional reconnection events occur on much shorter timescales. Between $t = 3.7~\mathrm{yr}$ (panel [a]) and $t = 4.0~\mathrm{yr}$ (panel [d]), the same averaged field line transitions from an extended state to a more compact (lower altitude) configuration. Given the continuous outflow during this interval, the contraction of the averaged field line suggests magnetic reconnection, which primarily occurs along the $B_r = 0$ boundary (highlighted by the magenta contour in Figure~\ref{fig:Br_flux}) where the open field lines threading the star and disk meet. Crucially, a full reset of the magnetic geometry into a closed dipolar loop is not required to sustain the jet: the required toroidal field can be generated by differential rotation as long as a star-anchored fieldline can become partially and shallowly embedded in the gas-dominated, magnetically elevated disk, as shown in sec.~\ref{sec:fire} and Fig.~\ref{fig:jet_twist_fieldlines}. Consequently, individual reconnection events occur more rapidly and on much shorter time scales than in 2D models, making the so-called ``magnetospheric ejections (MEs)" associated with prominent coherent reconnection events in 2D \citep[e.g.,][]{Zanni2013} harder to identify in 3D. How these 2D MEs are modified by severe tangling of magnetic field lines in the turbulent (MRI-active) and magnetically puffed-up inner disk remains to be quantified.

To summarize, jet launching operates through a ``load-fire-reload'' cycle driven by differential rotation between the star and disk. The twisting of magnetospheric field lines amplifies the toroidal magnetic field, generating a strong current sheet that accelerates the jet via the magnetic pressure force. Reconnection near the field reversal interface then resets the system, reforming field lines connecting the star and disk and reloading them with disk material for the next cycle. In our 3D simulation, individual reconnection events occur asynchronously across azimuth—that is, concurrently but without global coherence, with different $\phi$ sectors loading and releasing mass and magnetic energy independently. This spatial and temporal asynchrony sustains a rather continuous jet despite the intrinsically episodic nature of each local event.

\section{Discussion}
\label{sec:discussion}

\subsection{Magnetospheric accretion and outflow}
\label{subsec:MagnetosphericAccretion}

Although magnetospheric accretion is not the focus of this paper, we describe its general properties to set the stage for comparison with other jet-launching mechanisms.

Figure~\ref{fig:3D_magnetosphere} shows a 3D rendering of the accretion streams onto the star (the red sphere at the center) 
The accretion streams are filamentary and follow the magnetospheric field lines. These ``finger-like'' accretion patterns result from the ``interchange instability,'' a form of Rayleigh-Taylor instability in a magnetically dominated environment. Gas from the mass-dominated disk is loaded onto the strong magnetic field and subsequently accreted onto the star following the stellar magnetic field lines. This mechanism is established by theory \citep{Arons1976, Spruit1995} and simulations \citep[][]{Romanova2008, Blinova2016, Takasao2022, Zhu2024}.

These accretion streams originate from a mass-dominated disk whose boundary with the magnetically dominated magnetosphere marks the magnetospheric truncation radius $R_T$. We estimate the location of $R_T$ in two ways. 

First, we follow \citet{Zhu2025} and define the truncation radius analytically using the stellar and disk accretion parameters\footnote{ We note that there is a long history of analytically estimating the magnetospheric truncation radius, as discussed in, e.g., \citet{Bessolaz2008}. The analytic expression we adopted from \citet{Zhu2025} is identical to that recommended by \citet{Bessolaz2008} (see their equation [6]).}: 
\begin{equation}
    R_{T, \mathrm{ana}} = R_\star\Big(\frac{B_\star^4R_\star^5}{2GM_\star\dot{M}^2}\Big)^{1/7} = \Big(\frac{\bar{m}^4}{2GM_\star\dot{M}^2}\Big)^{1/7},
    \label{equ:Rt_ana}
\end{equation}
where $B_\star$ and $R_\star$ are the magnetic field strength on the stellar surface and radius of the star, respectively, $M_\star, \dot{M}$ are the stellar mass and mass accretion rate, respectively. $\bar{m}$ is the coefficient used in setting the strength of the stellar magnetosphere (equ.~\ref{equ:Amag}). Using our simulation parameters and $\dot{M} = 10^{-7}M_\odot~\mathrm{yr}^{-1}$ for the accretion rate at late times ($t\gtrsim 3$~yr) when the system settles to a quasi-steady state (see below, and Figure~\ref{fig:acc_theta_3Dstream}), the truncation radius $R_{T, \mathrm{ana}} \approx 0.088~\mathrm{au}$.

Second, we follow \citet{Takasao2022} and define $R_T$ as the smallest radius where $\beta_K = 1$. In our model, $R_{T, ~\beta_K}$ settles to a value $\approx 0.082~\mathrm{au} (\approx 5 R_\star)$ at late times, slightly lower than the analytical estimate. Since the accretion stream is filamentary (Figure~\ref{fig:3D_magnetosphere}), it is not surprising that the material with $\beta_K = 1$ can penetrate slightly into the magnetosphere. The truncation radius is comparable to the corotation radius $\approx 0.1~\mathrm{au}$. Our $R_T$ is larger than \citet{Takasao2022} compared to the stellar radius, since our stellar magnetic field is stronger than theirs. Using either definition, the truncation radius is smaller than the corotation radius, which is slightly larger than 0.1 au.

Most of the mass is accreted along magnetic field lines onto the star away from the midplane. To illustrate the landing locations of the gas on the star, we show in Figure~\ref{fig:acc_theta_3Dstream}(a) the averaged mass accretion rate per unit area at different polar angles  $\frac{\partial \dot{M}}{\partial A}(\theta)$ between 3.0~yr and 5.5~yr in our model measured right outside $r_\mathrm{fix}$ (sec.~\ref{sec:method}) at $0.02~\mathrm{au}$. 
Most accretion occurs around $\sim20^\circ$-$40^\circ$ from the polar axis, while the accretion around the midplane ($\theta=90^\circ$) is negligible. 

The highly filamentary high-altitude accretion does not imply a strong variation in the accretion rate. Figure~\ref{fig:acc_theta_3Dstream}(b) shows that throughout the later stage of the simulation between 3.0~yr and 5.5~yr, the total 
mass accretion rate remains around $1\times10^{-7}~\mathrm{M_\odot~yr^{-1}}$. In comparison, the jet outflow rate, defined as the total mass flux at $\pm 5~\mathrm{au}$ with $v_z^p > 10^7~\mathrm{cm~s^{-1}}$, is about 
$2\times10^{-9}M_\odot~\mathrm{yr^{-1}}$,
around $2\% \sim 5\%$ of the gas accretion rate.
The total mass outflow rate, measured at $\pm 5~\mathrm{au}$ with gas velocity $>10^6~\mathrm{cm~s^{-1}}$, is around $4\times10^{-8}M_\odot~\mathrm{yr^{-1}}$.
The total outflow rate is about $40\%$ of the total accretion rate, similar to the ratio in \citet{Takasao2022} and \citet{Zhu2024}, although the exact values differ mainly due to the difference in the chosen stellar magnetosphere magnetic field strength.

The outflow carries an angular momentum. To connect the angular momentum transported by the outflow with the disk Keplerian angular momentum, we define the ``Keplerian angular momentum radius'' following \citet{Tu2025b}
\begin{equation}
    R_a = \frac{1}{GM}\Big(\frac{\Phi_l}{\Phi_m}\Big)^2,
    \label{equ:keplerian_angmom_radius}    
\end{equation}
where $\Phi_l = \int \rho l v_z dA$ ($l$ is the specific angular momentum) and $\Phi_m = \int \rho v_z dA$ are the angular momentum flux and mass flux in the outflow, respectively. 
Figure~\ref{fig:acc_theta_3Dstream}(c) shows $R_a$ for the jet and the disk wind, respectively, measured at $z=5$~au. The jet $R_a\approx 0.6~\mathrm{au}$, much larger than its launch radius $\lesssim 0.2~\mathrm{au}$. This shows an abundance of angular momentum carried by the jet, initially stored in the electromagnetic form associated with the twisting of the field lines. 
The disk wind $R_a\approx 1.3~\mathrm{au}$, slightly larger than its launching radius $\lesssim 1~\mathrm{au}$. 

Both the jet and the disk wind carry significant angular momentum. In the following section (sec.~\ref{sec:compare_aas}), we will compare the accretion and outflow properties with those driven by the disk through the avalanche-accretion stream. Since the key distinction between disk-driven outflow and our model is the presence of a stellar magnetosphere, we will discuss its impact on the overall dynamics of the system.

\begin{figure*}
    \centering
    \includegraphics[width=\linewidth]{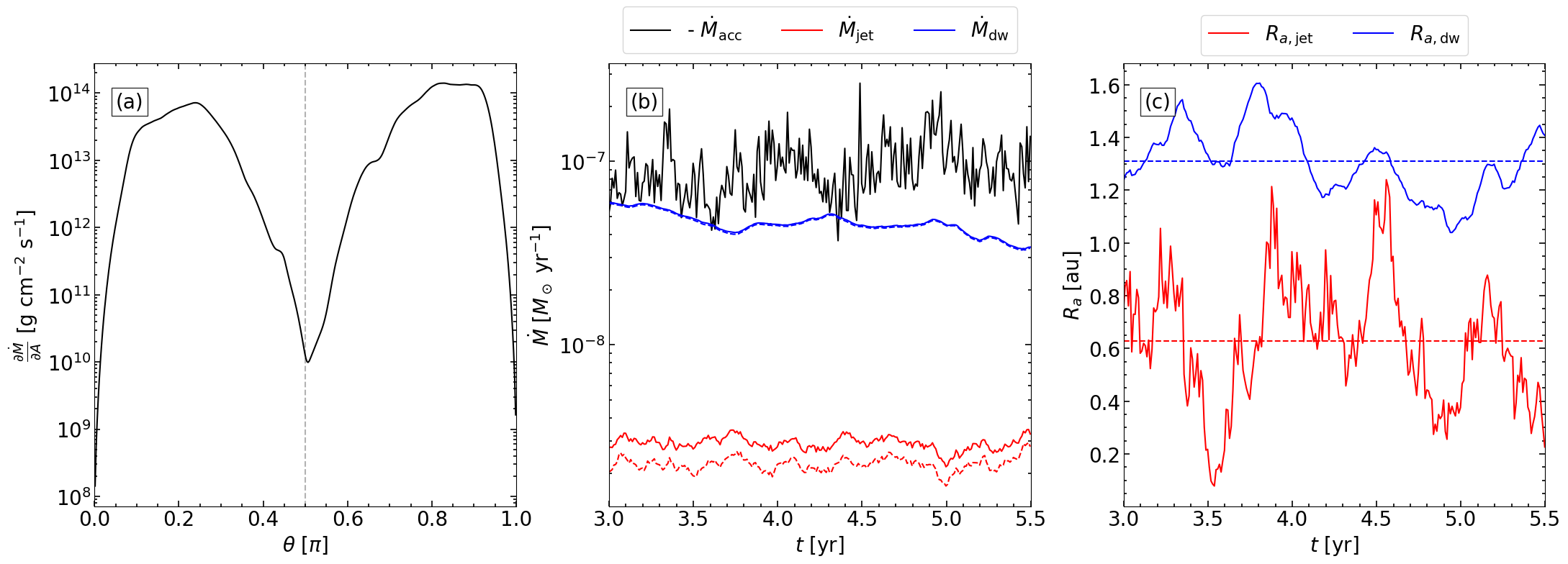}
    \caption{Accretion and outflow properties. \textbf{Panel (a)} quantifies the landing locations of the mass accretion rate streams on the star at each polar angle, showing off-midplane accretion along magnetosphere fieldlines. \textbf{Panel (b)} shows the total mass accretion rate (black line on top), jet outflow rate (red line at the bottom), and the disk wind outflow rate (blue line in the middle). The red dashed line is the jet outflow rate using the real gas only, showing the density floor's minimal effect on the outflow. \textbf{Panel (c)} shows the ``Keplerian angular momentum radius'' ($R_a$, equ.~\ref{equ:keplerian_angmom_radius}) of the jet (lower red line) and the disk wind (upper blue line). The horizontal dashed lines mark the averaged $R_a$ over the time shown.}
    \label{fig:acc_theta_3Dstream}
\end{figure*}

\subsection{Comparing to the disk-field-only jet model}
\label{sec:compare_aas}

An alternative jet-launching mechanism involves jets driven solely by the disk magnetic field. \citet{Tu2025a} and \citet{Tu2025b} model outflows in the absence of a stellar magnetosphere and identify a mechanism in which jets are powered by the so-called ``avalanche-accretion stream" in the magnetically elevated atmosphere of the disk. In this section, we compare the outflows driven by stellar magnetosphere-disk interactions with those produced by the disk magnetic field alone.  

To set the stage for this comparison, we first summarize the disk-field-only jet-launching mechanism described in \citet{Jacquemin-Ide2021}, \citet{Tu2025a}, \citet{Tu2025b}, and \citet{Mori2025}. In this scenario, the disk is threaded by a large-scale, open vertical magnetic field. Due to differential rotation and magnetic field geometry, the azimuthal magnetic field $B_\phi$ must switch sign somewhere between the upper and lower hemispheres, typically along the disk surface on one side. This sign reversal leads to strong azimuthal magnetic braking, which removes angular momentum and drives an avalanche-accretion stream along the disk surface. The differential rotation between this stream and the open field lines rapidly amplifies the toroidal magnetic field, generating strong toroidal magnetic pressure that lifts a portion of the accretion stream vertically, launching a jet. A similar process drives a slower, more mass-loaded disk wind at larger cylindrical radii, although this does not require the formation of an avalanche-accretion stream. Some regions of the inner disk wind can be magnetorotational instability (MRI)-active, where turbulent gas is lifted to high altitudes before merging with the disk wind.  

On the side of the disk without the azimuthal field sign change, the avalanche-accretion stream does not form, but jets can still emerge if the polar region remains relatively mass-free. However, MRI-active turbulent disk winds can fill the polar region with turbulent gas in many cases, overloading the outflow and effectively choking the jet. As a result, disk-field-only models frequently produce one-sided jets \citep[see also, e.g.,][]{Bethune2017, Takaishi2024}.  

A key distinction between the disk-only model and our magnetosphere-disk model is the persistence of a bipolar jet in our simulations (see Figure~\ref{fig:3D}). This difference arises from the global magnetic field geometry. The partial opening of the stellar magnetosphere introduces a substantial amount of open flux that threads the star in the polar regions. This open magnetic flux results in a strong field within the polar cavity, one that dominates over the disk magnetic field. As a result, the low-density polar cavity remains open in both hemispheres, allowing a stable bipolar jet to develop. In contrast, the disk-field-only model lacks this concentrated polar magnetic flux. Without the stellar contribution, the low-density jet cavity is more easily filled by the high-density disk-wind material, especially on the side where avalanche accretion does not occur. This configuration naturally leads to an asymmetric jet structure and, in some cases, a strongly one-sided jet. 

Another key distinction lies in how the toroidal magnetic field is generated. While both models rely on toroidal magnetic pressure to power the jet, the underlying mechanism for twisting the field differs significantly. In the disk field only model, the toroidal field is amplified by the differential rotation between the avalanche accretion stream of the disk surface and the slower-rotating gas above (and below) it. This configuration results in a toroidal twisting of the field where a faster-rotating layer below winds up a slower-rotating layer above.  

In contrast, the disk-magnetosphere model produces the toroidal field in an opposite manner. Here, strong magnetic field lines anchored to the rotating star transfer angular momentum to the (elevated) disk surface, spinning up the upper layers of the disk relative to the denser regions below. As a result, a toroidal field is generated by the differential rotation between the faster-rotating upper layers and slower-rotating lower layers of the disk. 

There are several important consequences of these differences in the generation of toroidal magnetic fields. One immediate effect is on the geometry of the poloidal magnetic field that channels the jet. In the disk-field-only model, the avalanche-accretion stream drags magnetic field lines inward, resulting in an inward-pinched poloidal field configuration (see Figure 5 in \citealt{Tu2025a}). In contrast, in the magnetosphere-disk model, the pinching point of the field moves outward, leading to an outward-pinched poloidal magnetic field structure (Figure~\ref{fig:jet_twist}).   

The difference in magnetic field geometry stems from differences in the power source of the jet. In the disk-field-only model, the disk itself provides all the energy and angular momentum for the outflow. In contrast, in the magnetosphere-disk model, the field lines anchored on the rotating star provide an additional source of energy and angular momentum to power the jet. In addition, once opened, the magnetosphere supplies the inner disk with a much stronger open poloidal field than in the disk-field-only case, and provides a global field configuration with oppositely directed open poloidal fields near the (elevated) disk surface that is conducive to magnetic reconnection and jet mass loading. As a result, jets in the magnetosphere-disk model are both more energetic and more massive than those in the disk-field-only model from \citet{Tu2025b}. Specifically, the jet in our simulations reaches velocities of several hundred km s$^{-1}$, compared to $\approx 100$ km s$^{-1}$ in the disk-field only case. The mass outflow rate is also higher, with $\dot{M}_\mathrm{jet} \approx 2\times10^{-9}~M_\odot\ \mathrm{yr}^{-1}$ compared to $\approx2\times10^{-11}~M_\odot\ \mathrm{yr}^{-1}$. Additionally, the jet carries more angular momentum, with the Keplerian angular momentum radius (equ.~\ref{equ:keplerian_angmom_radius}) of $R_a \approx 0.6~\mathrm{au}$, compared to $0.3~\mathrm{au}$ in the disk-field-only model.  

Similarly, the disk wind in the magnetosphere-disk model is stronger than in the disk-field-only model. This enhancement arises partly from the additional magnetic field strength in the disk due to the opening of the stellar magnetosphere. Moreover, some angular momentum in the disk wind originates from the star itself. When the magnetic field transfers energy and angular momentum to the jet, a portion of that angular momentum is fed into the disk.
Although this effect is relatively minor in terms of driving outflows (since the disk remains gas-dominated), it contributes to the overall angular momentum budget of the disk wind.  

In summary, the stellar magnetosphere plays two critical roles in shaping the outflow. 
First, the opening of the magnetosphere increases the field strength in the jet-launching inner disk region and imposes a global field geometry that is conducive to jet mass loading through reconnection and resistant to the choking of the jet by the disk wind. Second, by transferring energy and angular momentum from the rotating star to the outflow, the magnetosphere strengthens both the jet and the disk wind, making them more energetic and angular-momentum rich than in the disk-field-only scenario.  

\subsection{Magnetic-to-Kinetic Energy and Angular Momentum Conversion in the Jet} 
\label{sec:conversion}

There are four conserved quantities along magnetic field lines in the axisymmetric steady-state wind model (see, e.g., equation~32 to 35 in \citealt{Tu2025b}; also \citealt{Salmeron2011, Zhu2018, Jacquemin-Ide2021}). However, since our jet-launching mechanism (Sec.~\ref{sec:jet-launching-mechanism}) is highly variable, involving repeated opening and reconnection of field lines, these quantities are not strictly conserved along individual field lines. Nevertheless, we can still make a rough connection to three of them using the total fluxes of mass, poloidal field, angular momentum, and energy integrated over the cross-sectional area of the jet at different heights as follows. 

The first quantity 
\begin{equation}
    k' = \frac{\int_\mathrm{jet}4\pi\rho v_\mathrm{z}~dA}{\int_\mathrm{jet}B_\mathrm{z}~dA}
    \label{equ:cons_mass}
\end{equation}
is the ratio of the mass flux to the poloidal magnetic flux threading the jet. The second quantity 
\begin{equation}
    l'=\frac{\int_\mathrm{jet} \big[\rho v_\mathrm{z}(Rv_\phi) - \frac{R B_\phi B_\mathrm{z}}{4\pi}\big]~dA }{\int_\mathrm{jet} \rho v_\mathrm{z} dA}
    \label{equ:cons_angmom}
\end{equation}
is the angular momentum flux of the jet per unit mass flux. 
The third quantity 
\begin{equation}
    E'= \frac{\int_\mathrm{jet} \big[\frac{1}{2}\rho v^2 v_\mathrm{z} + \frac{ B^2 v_\mathrm{z}}{4\pi}-\frac{(\mathbf{B}\cdot\mathbf{v})B_\mathrm{z}}{4\pi}\big] dA}{\int_\mathrm{jet}\rho v_\mathrm{z}dA}
    \label{equ:cons_ene}
\end{equation}
is the total (kinetic and magnetic) energy flux per unit mass flux. Both thermal and gravitational terms are ignored since they are negligible compared to the kinetic and magnetic terms. 

\begin{figure*}
    \centering
    \includegraphics[width=0.8\linewidth]{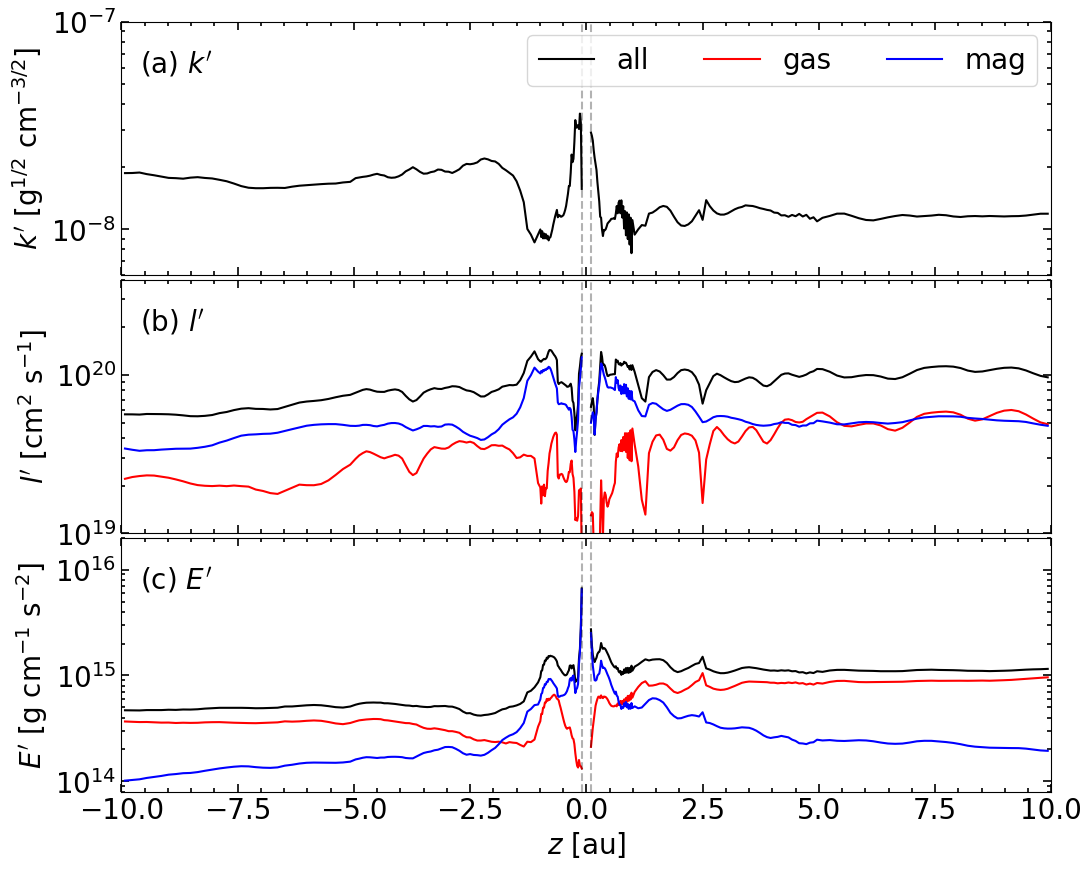}
    \caption{Integrated jet quantities defined in equations \ref{equ:cons_mass}-\ref{equ:cons_ene} as a function of height $z$: (a) the mass flux to magnetic flux ratio, (b) the angular momentum flux per unit mass flux, and (c) the energy flux per unit mass flux, showing the magnetic domination of the angular momentum and energy fluxes near the jet base and conversion of magnetic angular momentum and energy to the gas at larger distances.}
    \label{fig:cons_3D_asfunz}
\end{figure*}

Figure~\ref{fig:cons_3D_asfunz}(a) shows $k'$, which is roughly constant along the jet. This indicates that the mass loading of the outflow remains approximately the same as the jet propagates outward. Figure~\ref{fig:cons_3D_asfunz}(b) and (c) show $l'$, and $E'$, and their gas and magnetic component, respectively. Near the jet base ($\lesssim 2~\mathrm{au}$), the magnetic field dominates the transport of both angular momentum and energy. Farther out, the angular momentum carried by the gas becomes comparable to that of the magnetic field, while the energy flux $E'$ is increasingly dominated by the gas component. This transition suggests an ongoing transfer of energy and angular momentum from the magnetic field to the gas as the outflow expands, leading to a progressively gas-dominated jet at larger scales. The total angular momentum carried by the jet is substantial, especially when the magnetic contribution is included, which exceeds the gas-only angular momentum estimated in Sec.~\ref{subsec:MagnetosphericAccretion}. These findings are consistent with, and similar to, the analysis along the field line in \citet{Tu2025a, Tu2025b, Jacquemin-Ide2021}.

\vskip 0.5cm

\subsection{ Towards a Scenario of Accretion-fed Polar Poynting Jet Driven Along Open Stellar Field Lines Through Disk-Magnetosphere Interaction}

\begin{figure*}
    \centering
    \includegraphics[width=\linewidth]{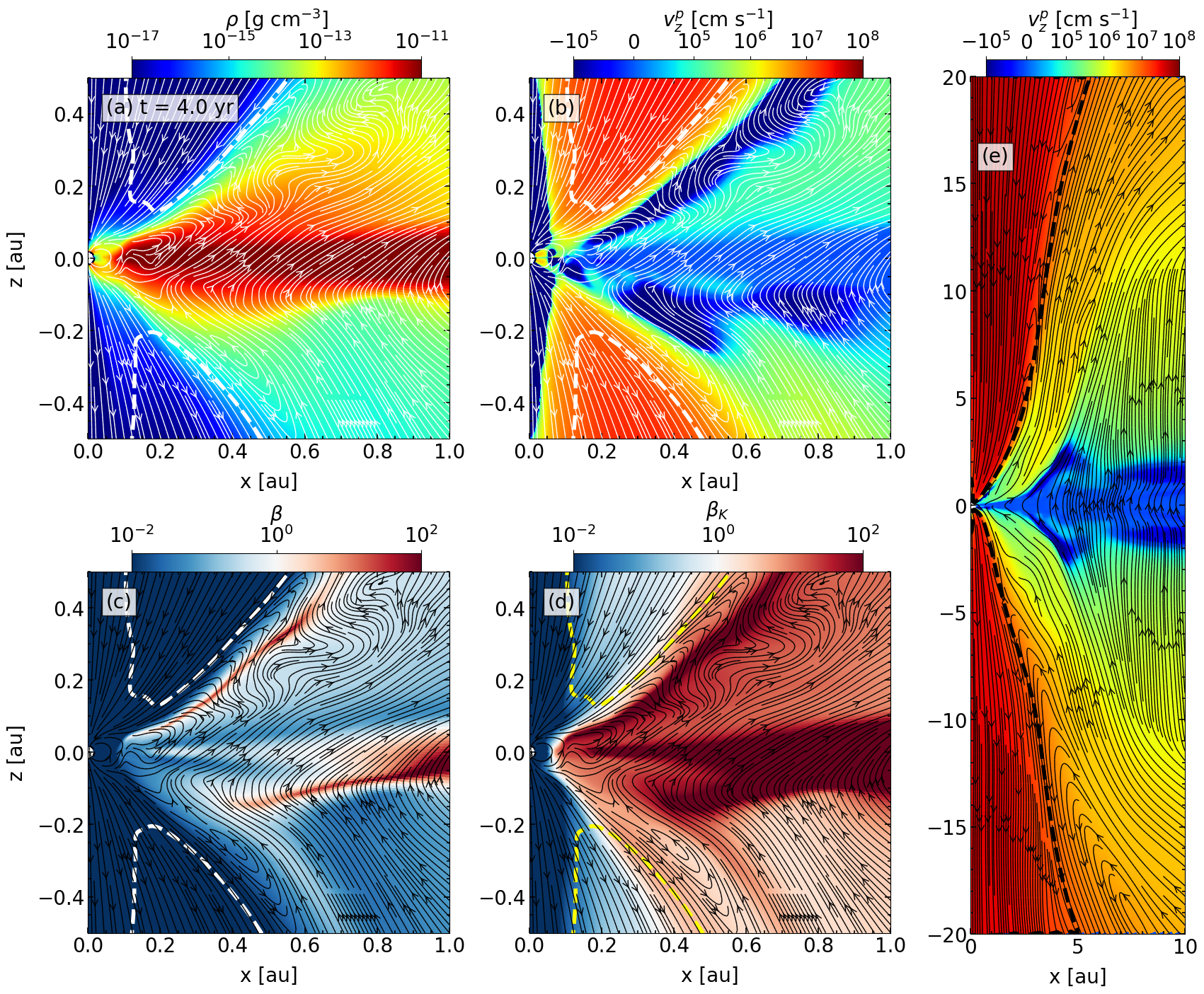}
    \caption{Flow quantities averaged azimuthally and over time (from 3.5 to 4.5~years) to highlight the differences between the low-density magnetically dominated fast polar jet and its surrounding, denser, matter-dominated more slowly moving disk atmosphere/disk wind in (a) density, (b) vertical velocity away from the midplane, (c) the regular (thermal) plasma-$\beta$, and (d) the kinetic plasma-$\beta_K$ on a relatively small scale, and (e) vertical velocity away from the midplane on a larger scale. Magnetic field lines are plotted in each panel, showing that the low-density, magnetically dominated, fast-moving jet (with $v_\mathrm{z}^p > 10^7$cm/s) flows outward primarily along open stellar field lines. }
    \label{fig:StellarWind}
\end{figure*}

One of the most striking features of our simulation is that the fast, collimated jet material appears to flow primarily along open magnetic field lines that are anchored on the central star. It can already be seen from the azimuthally averaged flow quantities plotted in Fig.~\ref{fig:jet_twist}[b] (and its associated animation; see also Fig.~\ref{fig:diskwind}[b] on a larger scale), and is illustrated more clearly in Fig.~\ref{fig:StellarWind}, where we show flow quantities averaged both azimuthally and over time (from 3.5 to 4.5~yrs or about 30 stellar spin periods). 

The magnetic field configuration in Fig.~\ref{fig:StellarWind} is similar to that in Fig.~1 of \citet{Zanni2013}, which displayed the time-averaged magnetic field and flow patterns from their reference 2D (axisymmetric) disk-magnetosphere interaction simulation with a prescribed magnetic diffusivity and fluid viscosity. They found four distinctive regions: (1) a stellar wind, where a fast outflow is driven thermally along open stellar field lines, (2) a disk wind, where a slower outflow is driven by field lines anchored on the disk, (3) a magnetospheric ejection (ME) region between the disk wind and stellar wind, where field lines connecting both the star and the disk are episodically inflated and reconnected, ejecting plasmoids with detached field lines, and (4) a region where the stellar magnetic field is more persistently connected to the disk. 

Since our fast-moving jet primarily flows out along open stellar field lines, it would be called a stellar wind in Zanni and Ferreira's terminology. However, there is a crucial difference between our jet and their stellar wind. The material in their stellar wind comes from the stellar surface, accelerated outwards by a prescribed thermal pressure along the open stellar field lines. In contrast, no hot gas capable of escaping to large distances is prescribed near the stellar surface in our simulation; as a result, the gas near the stellar surface is pulled by strong gravity inward toward the star along the open field lines in the polar region rather than flowing outward (see Fig.~\ref{fig:StellarWind}[b]). The outflowing material in our simulation appears to be loaded onto the open stellar field lines well above (and away from) the stellar surface, likely through magnetic reconnection, which is expected to be more chaotic and localized in our 3D global simulation without a prescribed magnetic diffusivity and fluid viscosity (as discussed below), although numerical diffusion may also contribute. Determining the exact origin of the jet mass loading is a pressing issue that needs to be addressed in future investigations, such as through higher resolution simulations to reduce numerical diffusion.

Although the jet material ultimately flows along open stellar magnetic field lines, these lines are not necessarily attached {\it only} to the star {\it at all times}. They typically undergo a cyclic evolution in which a given field line is first connected simultaneously to the star and the disk (or its elevated, turbulent atmosphere), allowing dense disk gas and magnetic stress to be transferred onto it. Magnetic reconnection then severs the disk footpoint, leaving the line open and anchored only on the star; the toroidal magnetic pressure built up during the connected phase is released, accelerating the newly loaded plasma through the low-density polar cavity (which is kept open by the star-anchored, predominantly open field lines in the cavity in the first place). In this way, the jet is instantaneously a star-anchored outflow, but is powered and mass-loaded by earlier episodes in which the now-open field lines interact with the denser disk/atmosphere.  This transient two-footpoint phase is essential for understanding how a magnetically driven, star-anchored Poynting jet can remain continuously energized and supplied with mass.

The differences between our work and \citet{Zanni2013} in the level of (explicit) magnetic diffusivity and in the simulation's dimensionality have important consequences. First, the absence of (explicit) magnetic diffusion enables rapid winding of the initially weak magnetic field in the inner disk, leading to a buildup of magnetic pressure that forms a highly elevated, dynamic (indeed MRI-unstable) disk atmosphere, which gradually merges into a slow, dense disk-wind. The puff-up facilitates a more continuous jet driving because the open stellar field lines (which have a natural tendency to curve back toward the midplane and return to the initial dipolar configuration) only need to touch, or shallowly embedded in, the elevated disk atmosphere (as illustrated by the green and magenta field lines in Fig.~\ref{fig:jet_twist}) to drive outflow. These ``touch-and-go'' field lines, which are a subset of the ``two-legged'' field lines, increase the rapidness of mass ejection by {\it not} requiring a full reset of the global magnetic field geometry. Just as importantly, the extra dimension in our 3D simulation enables magnetic reconnection to occur asynchronously across azimuth—concurrently but without global coherence—so that different $\phi$ sectors load and release mass and magnetic energy independently. It makes the local reconnection-driven ``load-fire-reload" jet-launching cycle much shorter than those observed in 2D simulations, which can last for several stellar rotation periods \citep[e.g.,][]{Zanni2013}. The more frequent and localized reconnection events enabled by the third (azimuthal) dimension and the disk puff-up alleviate the need for the global large-scale episodic outbursts often seen in 2D simulations. They facilitate the rather continuous, quasi-steady mass ejection seen in our jet (see the animated version of Fig.~\ref{fig:3D}), particularly at later times after the initial adjustments. The suppression of large-scale temporal variations in 3D is consistent with the findings of \citet{Takasao2022}. 

Our polar jet can be viewed as a variant of the so-called ``accretion-powered stellar wind" advocated by \citet[][compare our Fig.~\ref{fig:StellarWind}a,b with the cartoon sketched in their Fig.~1]{Matt2005A} in the context of braking the stellar spin. They envisioned several processes for transferring some of the accretion energy to the corona region of open stellar field lines, such as accretion shocks, magnetic reconnection, and magnetic waves, which may raise the stellar corona's temperature and thermally drive a stellar wind, as simulated by \citet{Zanni2013} and others \citep[e.g.,][]{Matt2008}. However, our jet has a thermal pressure orders of magnitude smaller than the magnetic pressure (see Fig.~\ref{fig:StellarWind}[c]) and is thus completely magnetically driven; it is essentially an electromagnetic (Poynting) jet powered by field line bending through the interaction between the stellar field lines and the surrounding disk (and its elevated atmosphere and wind), with the Poynting energy flux gradually converted into the kinetic energy flux as the jet propagates outward (see Fig.~\ref{fig:cons_3D_asfunz}[c]). 

The Poynting jet in our simulation has more in common with the magnetically dominated axial jet found in the 2D (axisymmetric) simulations of \citet[][see also \citealt{Romanova2009}]{Ustyugova2006} in the strong ``propeller" regime of disk-magnetosphere interaction than with the thermally driven stellar wind. One major difference is that they prescribed an explicit magnetic diffusivity and fluid viscosity, as in \citet{Zanni2013}. The explicit magnetic diffusivity allows the material flowing along the upper funnel flow region of the closed magnetosphere to diffuse to its neighboring open stellar field lines (see, e.g., Fig.~11d of \citealt{Ustyugova2006} or Fig.~19 of \citealt{Romanova2009}), thereby providing an elegant explanation for the jet mass loading. Our 3D simulation shows that mass loading onto open stellar field lines is possible even without explicit magnetic diffusivity, likely due to a more chaotic flow in both space and time, as mentioned earlier, which results in an effective cross-field mass diffusion, plausibly between the open stellar field and both the closed magnetospheric field carrying funnel flows and the field threading the elevated disk atmosphere/disk-wind bordering the jet. 

The explicit magnetic diffusivity and viscosity in \citet{Ustyugova2006} is also likely the reason why their rather laminar accretion flow can cross well-ordered poloidal field lines near the inner edge of the disk and be ejected as a conic wind surrounding the axial jet (see their Fig.~11).  In contrast, the absence of an explicit magnetic diffusivity in our simulation facilitates field line winding in and around the inner disk through differential rotation, which puffs up the disk atmosphere and drives a broader disk-wind surrounding the axial jet (see Fig.~\ref{fig:Br_flux}[a]). 

Another difference is that \citet{Ustyugova2006} only found strong jets in the strong ``propeller" regime.  In our simulation, the truncation radius is comparable but slightly smaller than the corotation radius, placing the system in the non-propeller or, at most, weak propeller regime. However, a strong jet is still produced. One plausible way to reconcile this apparent contradiction is that the stellar field lines in our simulation are connected to the elevated disk atmosphere at higher heights and larger radii, where the gas rotates more slowly than at the inner disk edge on the midplane, creating an ``effective propeller" regime above and below the disk that enables energy and angular momentum to be extracted magnetically from the rotating star to power the jet. 

Nevertheless, in both our simulation and those of \citet{Ustyugova2006}, a fast collimated magnetically dominated axial jet is produced despite significant differences in magnetic diffusivity, fluid viscosity, simulation dimensionality (2D vs 3D), and whether the system is in the strong ``propeller" regime. This consistency in outcome points to a robust mechanism for producing the accretion-fed, tenuous axial Poynting jet driven in the low-density polar cavity along open stellar magnetic field lines by field line bending through the magnetosphere-disk interaction. Crucially, the interaction-opened stellar magnetic field pushes against the elevated disk atmosphere/disk wind to keep the low-density polar cavity open, enabling the light mass-loading critical for the jet acceleration to a high speed.

This lightly loaded, magnetically dominated, axial jet is physically distinct from the traditional disk wind launched over a range of disk radii or the X-wind launched near the inner edge of the disk. In particular, it is {\it not} the densest part of a magneto-centrifugal disk- or X-wind that extends all the way to the axis. A disk- and/or X-wind can still exist outside the axial Poynting jet. They would have the same general source of mass loading (accreting material either in the disk/atmosphere or funnel flow), but they would be launched by field lines anchored primarily on the disk rather than the star. As such, unlike the disk- or X-wind, the jet in our simulation can extract angular momentum from the star directly, although we refrain from a quantitative analysis of the stellar braking by the jet because of the uncertainties about the exact details of its mass loading.   

It is possible that the part of the disk- and/or X-wind immediately surrounding the axial Poynting jet can be collimated toward the axis, forming an outer, more mass-loaded, layer of the global jet system at large distances. Furthermore, it is likely that the central YSO blows a tenuous (Sun-like) stellar wind, which may form yet another layer (likely the innermost and fastest) of the jet (see \citealt{Ferreira2006} for a discussion)\footnote{This stellar wind would likely prevent the infall in the polar region in our simulation (see, e.g., the left panel of Fig.~\ref{fig:init_open}), although its effects on the global 3D flow structure remain to be quantified.}. The existence of different jet components is hinted at, for example, by the detection of an X-ray jet in DG Tau, which indicates a faster speed than measured in the optical/IR component of the jet (as sketched in, e.g., Fig.~3 of \citealt{Gunther2009}. In any case, the existence of a Poynting jet as described in this work and that of Usytugova et al. does not preclude other jet components of different origins. Future investigations should focus on the relative importance of these potential components under different conditions and on their possible interactions.

\section{Conclusions}
\label{sec:conclusions}

We have carried out global, three-dimensional magnetohydrodynamic simulations of the interaction between a rotating, magnetized young star and its accretion disk, focusing on how the coupled system launches and sustains an axial jet. The simulation follows the evolution of both accretion and outflow without prescribing magnetic diffusivity or viscosity, allowing magnetic reconnection, mass loading, and jet launching to emerge more self-consistently. Our main findings are summarized below.

\begin{enumerate}
    \item Our 3D simulation reproduces the essential \textit{load--fire--reload} magnetic cycle identified in earlier 2D studies (e.g., \citealt{Zanni2013}) but extends it in several crucial ways. In the absence of ad hoc magnetic diffusivity or viscosity, the system develops a magnetically elevated, MRI-active inner disk that feeds mass onto open, star-anchored field lines via localized reconnection events. The third (azimuthal) dimension enables the cycle of loading, firing, and reloading to occur rapidly and asynchronously across azimuth---concurrently but without global coherence---producing a quasi-steady, continuously replenished jet rather than the large, periodic ejections typical of axisymmetric models. Thus, a fast, magnetically dominated jet can arise self-consistently from turbulent disk--magnetosphere coupling in ideal MHD.

    \item The physical picture that emerges is that of an \textit{accretion-fed, star-anchored Poynting jet} operating in the low-density polar cavity. Star-anchored field lines extract rotational energy from the star and inner disk, periodically reconnecting with the magnetized disk and its elevated atmosphere to reload mass and magnetic stress. Each field line cycles between a two-footpoint phase, where mass and twist accumulate, and a one-footpoint phase, where magnetic pressure accelerates plasma outward. Differential rotation regenerates the toroidal field, and asynchronous reconnection across azimuth maintains a continuous, large-scale jet. The jet in this scenario is lightly mass-loaded, as in the classical magneto-centrifugal model, but accelerated by magnetic pressure gradients, as in the magnetic-tower model. Its formation is facilitated by the low-density polar cavity kept open by star-anchored open field lines. 

    \item The global outflow consists of two coupled components: a \textit{fast, lightly loaded Poynting jet} confined to the polar cavity and a \textit{slower, denser disk wind}. The jet is not the densest axial part of the disk wind, which, in our simulation,
    is launched through the classical magnetic-tower mechanism. The denser disk wind dominates the system's mass loss and provides a confining sheath for the jet, while the jet itself contains a significant toroidal magnetic field that contributes to self-collimation. 

    \item The stellar magnetosphere provides the star-anchored open field lines that maintain the bipolar cavities (one in each hemisphere), ensuring bipolar jets and preventing choking by turbulent disk material. In contrast to disk-only models that often produce one-sided or intermittent outflows, the inclusion of the rotating stellar field yields a persistent, bipolar structure in which the jet extracts energy and angular momentum from both the star and the disk. The appearance of similar fast, magnetically dominated jets across simulations with different microphysical assumptions---ranging from magnetically diffusive 2D (e.g., \citealt{Ustyugova2006}) to ideal MHD 3D---indicates that \textit{magnetic-pressure acceleration of the accretion-fed low-density material in the polar cavity to high speed through disk--magnetosphere interaction} is a robust mechanism for jet production in young stellar systems. Higher-resolution simulations are desirable to firm up this conclusion.
\end{enumerate}

\begin{acknowledgments}

We thank the referee for detailed and constructive comments that significantly improved the paper. YT acknowledges support from the interdisciplinary fellowship at UVA. CYH acknowledges support from an ALMA Student Observing Support (SOS) from NRAO. We acknowledge computing resources from UVA research computing (RIVANNA), NASA High-Performance Computing, and NSF's ACCESS computing allocation AST200032 and PHY250098. ZYL, YT, and CYH are supported in part by NASA 80NSSC20K0533, NSF AST-2307199, JWST-GO-02104.002-A, and the Virginia Institute of Theoretical Astronomy (VITA). Z.Z. acknowledges support from NSF award 2408207, 2429732 and NASA grant 80NSSC24K1285.

\end{acknowledgments}

\begin{contribution}
YT led the modeling, analysis, interpretation, and writing of the paper. ZYL and Z.Z. contributed to the analysis and interpretation, and oversaw the project's progress. HX and CYH contributed to the interpretation.


\end{contribution}

\bibliography{sample7}{}
\bibliographystyle{aasjournalv7}



\end{document}